\documentclass[11pt]{article}
\pdfoutput=1
\usepackage{amsmath,amssymb,amsfonts}
\usepackage[nosort]{cite}
\usepackage{graphicx}
\usepackage[colorlinks=true, pdfstartview=FitV, linkcolor=blue, citecolor=magenta, urlcolor=purple]{hyperref}
\usepackage{mdframed,framed}
\usepackage{cancel}
\usepackage[usenames,dvipsnames]{xcolor}
\usepackage{notoccite}
\newcommand{\beq}{\begin{equation}}
\newcommand{\eeq}{\end{equation}}
\newcommand{\beqn}{\begin{eqnarray}}
\newcommand{\eeqn}{\end{eqnarray}}
\newcommand{\pa}{\partial}

\usepackage{cancel}

\newcommand\Lie{\pounds}

\newcommand{\un}[1]{\underline{#1}}

\newcommand{\comment}[1]{}

\newcommand{\D}{\text{d}}

\usepackage{databib}
\newcommand{\bs}{\pmb{\sigma}}
\newcommand{\be}{\pmb{\epsilon}}
\newcommand{\bx}{\pmb{\xi}}
\usepackage{tocloft}
\usepackage{indentfirst} 
\numberwithin{equation}{section}
\newcommand{\thistitle}{Boundary Conditions for Warped AdS$_3$ in Quadratic Ensemble}
\newcommand{\ulb}[1]{
	\centerline{
		\begin{minipage}[c]{0.7\textwidth}
			\begin{center}
			${}^{#1}$ Physique Math\'ematique des Interactions Fondamentales \& International Solvay Institutes, Universit\'e Libre de Bruxelles, Campus Plaine - CP 231, 1050 Bruxelles, Belgium\\
			\vspace{0.5cm}
\href{mailto:Ankit.Aggarwal@ulb.be}{Ankit.Aggarwal@ulb.be}, \hspace{1 cm}
		\href{mailto:Luca.Ciambelli@ulb.be}{Luca.Ciambelli@ulb.be},
\hspace{1.1 in}
			\href{mailto:Stephane.Detournay@ulb.be}{Stephane.Detournay@ulb.be},   \hspace{.2cm}
		\href{mailto:Antoine.Somerhausen@ulb.be	}{Antoine.Somerhausen@ulb.be}
			\end{center}
		\end{minipage}
		}
	}
\newcommand{\ams}[1]{
	\centerline{
		\begin{minipage}[c]{0.7\textwidth}
			\begin{center}
			${}^{#1}$ Institute for Theoretical Physics Amsterdam and Delta Institute for Theoretical \\ Physics, University of Amsterdam, Science Park 904, 1098 XH \\ Amsterdam, The Netherlands
			\end{center}
		\end{minipage}
		}
	}

\begin{document}
\title{\thistitle}
\author{
	Ankit Aggarwal,$^{a,b}$ Luca Ciambelli,$^{b}$ St\'ephane Detournay$^{b}$ and Antoine Somerhausen$^{b}$
	\\
	\\
	\\
	{\small \emph{\ams{a}}} \\ \\
	{\small \emph{\ulb{b}}}
	\\
	\\
	}
\date{}
\maketitle
\vspace{-5ex}
\begin{abstract}
\vspace{0.6cm}
In the context of warped conformal field theories (WCFT), the derivation of the warped Cardy formula relies on the zero mode spectrum being bounded from below. Generically, this is not true for holographic WCFTs in ``canonical'' ensemble, whereas this condition is satisfied in the ``quadratic'' ensemble, making it more natural in holographic setups.
In this work, we find a new set of boundary conditions in three-dimensional Topologically Massive Gravity (TMG) such that the putative dual theory is a WCFT in quadratic ensemble. Surprisingly, imposing the equations of motion, we obtain a closed form metric spanned by  two arbitrary chiral functions, analogous to the Ba\~nados metrics in Einstein gravity. Surface charges for these boundary conditions are not a priori integrable and we discuss  two choices of boundary conditions to make them so. We obtain the bulk thermodynamic entropy of warped BTZ black holes by making use of the warped Cardy formula, in its regime of validity. We also discuss the issue of identifying the vacuum in our solution space:   demanding the enhancement of global symmetries selects only a family of solutions, out of which the unique vacuum must be carefully singled out.
\end{abstract}

\newpage

\begingroup
\hypersetup{linkcolor=black}
\tableofcontents
\endgroup
\noindent\rule{\textwidth}{0.8pt}

\newpage

\section{Introduction}
\indent
In the search for toy models able to capture the main features of the most salient black hole conundrums, lower-dimensional theories \cite{Teitelboim:1983ux, Deser:1983tn, Deser:1983nh, Jackiw:1984je} have always stood out.\footnote{The literature, especially for $3$-dimensional gravity, is vast. Some standard useful books and reviews are \cite{doi:10.1142/0622, doi:10.1142/2295, Carlip:1995zj}.} The seminal work of Brown and Henneaux \cite{Brown:1986nw} has for the first time pointed at a connection between AdS gravity and a CFT, and BTZ black holes \cite{Banados:1992wn, Banados:1992gq} have successfully been used in a variety of contexts to explore properties of gravity beyond the classical regime. In recent years, the principles of AdS/CFT have been applied to Kerr black holes with potential groundbreaking applications in astronomy \cite{Guica:2008mu}. The status of the correspondence is however not settled as several consistent and possibly related proposals exist \cite{El-Showk:2011euy, Compere:2012jk, Detournay:2012pc, Guica:2017lia, Haco:2018ske, Aggarwal:2019iay}. 

A useful lower-dimensional toy model for Kerr black holes has emerged over the years in the form of a deformed BTZ spacetime, the so-called Warped AdS$_3$ (WAdS$_3$) black holes, \cite{Rooman:1998xf, Duff:1998cr, Israel:2003ry, Israel:2004vv,  Bouchareb:2007yx, Carlip:2008eq, Anninos:2008fx, Compere:2008cv, Henneaux:2010fy}.  WAdS$_3$ black holes display features departing from AdS$_3$ black holes and closer quantitatively to Kerr. For instance, the near-horizon geometry of extremal black holes has $SL(2,R) \times U(1)$ isometry like the near horizon region of extremal Kerr black holes (NHEK) \cite{Bardeen:1999px}; actually, it is a section of constant polar angle of NHEK. Also, their Penrose diagram resembles that of an asymptotically flat black hole \cite{Jugeau:2010nq}.
 The study of the asymptotic symmetry group of WAdS$_3$ black holes in the spirit of Brown and Henneaux has revealed that they belong to a phase space with symmetries consisting in the semi-direct product of a chiral Virasoro algebra with an affine u(1) current \cite{Anninos:2008fx, Compere:2009zj, Blagojevic:2009ek, Henneaux:2011hv}. This suggested the existence of a new class of 2d field theories, Warped Conformal Field Theories (WCFTs) \cite{Detournay:2012pc}, invariant under translations and chiral scaling but not rotations \cite{Hofman:2011zj}, to which WAdS$_3$ would be dual to. The symmetries of WCFT are, to some extend, almost as powerful as a $2$-dimensional CFT. For this reason, many of the questions that can be explored and answered in a $2$-dimensional CFT have also become a subject of interest for WCFTs in the last few years: bootstrap techniques \cite{Apolo:2018eky}, the study of partition functions \cite{Detournay:2012pc, Castro:2015uaa}, and their entanglement entropy \cite{Castro:2015csg, Song:2016gtd, Song:2016pwx}. 
Interestingly, examples of WCFTs have been found in  chiral Liouville gravity \cite{Compere:2013aya}, free Weyl fermions \cite{Hofman:2014loa}, and free scalars \cite{Jensen:2017tnb}.

States in a WCFT can be labelled by the zero modes of the algebra (\ref{CSA}-\ref{CSAf}): $L_0$ and $P_0$. However, we are free to use another combination of generators to refer to them, for instance, the zero modes of the algebra (\ref{QSA}-\ref{QSAf}), $\tilde{L}_0$ and $\tilde{P}_0$.  
From these generators, one can build two different density matrices, $\rho_c = e^{-\beta P_0 +i \theta L_0}$ corresponding to what we refer to as {\itshape canonical} ensemble
 and $\rho_q = e^{-\beta_L \tilde{P_0} - \beta_R \tilde{L_0}}$ corresponding to what we will call {\itshape quadratic} ensemble. The entropy of a thermal state is independent of the ensemble. Now, for $\rho_c$, which is the ensemble usually considered for WAdS$_3$ black holes, we cannot use a Cardy like formula to derive the black hole entropy, because the conditions to derive this Cardy-like formula are not satisfied by the dual bulk spacetime, because the spectrum of $L_0$ is not bounded from below. For $\rho_q$ instead, we do know how to proceed, and we can compute the entropy. The charges $(L_n,P_n)$ (called ``canonical") and  $(\tilde{L}_n, \tilde{P}_n$) (dubbed ``quadratic")  are related to one another, and yield slightly different algebras (referred to as canonical \eqref{CSA}-\eqref{CSAf} and quadratic algebras \eqref{QSA}-\eqref{QSAf}). This operation at the boundary has a counterpart in the bulk: it is a non-local change of coordinates, which amounts to two different choices of boundary conditions. Performing it on the regular WAdS$_3$ black hole solution (dual to $\rho_c$), one gets the black hole solution dual to $\rho_q$ (Warped BTZ). The goal of this paper is to write boundary conditions including these solutions, and to have a bulk realization of the boundary relation between canonical and quadratic generators. For the zero mode charges, and solutions, this is trivial and spelled out in \cite{Detournay:2012pc}. We here extended this to the whole phase space and asymptotic Killing vectors. The phase space obtained is shown to have a symmetry algebra of WCFT expressed in quadratic charges, as expected. 
 The boundary conditions we obtain are a natural generalization of the Comp\`ere, Song, and Strominger (CSS) boundary conditions \cite{Compere:2013bya} for AdS$_3$ spaces (which in our terminology are in quadratic ensemble, satisfying (\ref{QSA}-\ref{QSAf})), to allow for more leading metric components as one approaches the boundary.\footnote{In that sense, the boundary conditions of \cite{Compere:2009zj} for $\mu\ell = \pm3$, i.e. in the AdS$_3$ limit, are the CSS boundary conditions in canonical ensemble described in App. B of \cite{Compere:2013bya}.}

 The paper is organized as follows. In Sec. \ref{s2}, we review WAdS$_3$ black holes and the bulk change of coordinates to go to the WBTZ line element. We discuss how WAdS$_3$ black holes are dual to a WCFT in canonical ensemble while the WBTZ line element is dual to the quadratic ensemble. We further show that the warped Cardy formula is well-defined only for WCFTs dual to WBTZ coordinates. This serves as a motivation for introducing in Sec. \ref{s3} new boundary conditions naturally encompassing WBTZ black holes. We further show in this section two relevant limits of our new solution space: the null warped and CSS limit. A discussion of the various regimes as a function of the TMG Chern-Simons coupling constant is also presented. In Sec. \ref{s4}, we compute the surface charges in TMG, and find that they are not integrable in general. We propose two ways to render them integrable: a restriction of the solution space, giving rise to a solution dual to a thermal state in a WCFT in quadratic ensemble; and a redefinition of the symmetry generators, giving rise to a solution dual to a thermal state in a WCFT in canonical ensemble. We subsequently perform the entropy analysis in Sec. \ref{s5}. We compute the bulk thermodynamic entropy and compare it with the WCFT Cardy formula, showing that they match once the vacuum is correctly identify. This is a delicate procedure because the standard enhancement of isometries  is not enough to single out the vacuum. This fact opens the door to interesting future investigations, that we discuss in the conclusions, after summarizing the main achievements of the paper. We exile to Appendix \ref{AppA} the study of new boundary conditions that include pp waves. 

\section{From WAdS$_3$ to Warped BTZ and warped Cardy formula}\label{s2}

We present in this section a self-contained review of the WAdS$_3$ analysis, the bulk change of coordinates to WBTZ black holes, and the warped Cardy formula which is counting the degeneracy of states in the dual WCFT. We detail the conditions on the spectrum to ensure the well-definiteness of this formula, and show that these conditions are met in holographic setup in the quadratic ensemble, but not in the canonical ensemble.

The line element for Warped black holes is (see e.g. \cite{Israel:2004vv, Bouchareb:2007yx, Carlip:2008eq, Anninos:2008fx})
\beqn\label{WBH}
\D s^2 =\frac{\text{d$\rho $}^2}{\frac{12 j}{\mu }+\frac{1}{9} {\rho ^2\over \ell^2} \left(\mu
   ^2\ell^2+27\right)-12 m \rho }+ \ell^2\text{dT}^2-\frac{4}{3}
     \mu \ell \rho  \ \text{dT} \ \text{d$\psi $}+
   \left({\rho^2\over 3 \ell^2}\big( \mu^2 \ell^2 -3\big)+12
   m\rho-\frac{12 j}{\mu }\right)\text{d$\psi $}^2
\eeqn
with $\rho\in [0,\infty)$, $T\in (-\infty,\infty)$, and $\psi\sim \psi+2\pi$. This solves the $3$-dimensional Topologically Massive Gravity (TMG) \cite{DESER1982372} equations of motion
\beq
R_{\mu\nu}-{R\over 2}g_{\mu\nu}+\Lambda g_{\mu\nu}+{1\over\mu}C_{\mu\nu}=0.\label{TMGeom}
\eeq
This theory depends on two parameters: the cosmological constant $\Lambda$ and the Chern-Simons coupling $\mu$. While the equations of motion \eqref{TMGeom} have solutions for all values of $\mu$, we can, without loss of generality, assume $\mu>0$ in the following. This comes about because each negative-$\mu$ solution can be rewritten with $\mu>0$ sending $\mu\to -\mu$, $\psi\to -\psi$, and $j\to -j$. Indeed, due to the conventions in the Cotton Hodge dual, defined below, eq. \eqref{TMGeom} are invariant under this symmetry. The Cotton Hodge dual is defined as
\beq
C_{\mu\nu}=\eta_{\mu}{}^{\rho\sigma}\nabla_\rho(R_{\sigma\nu}-{1\over 4}R g_{\sigma\nu}) \qquad \text{with} \qquad \eta_{\mu\nu\rho}=\sqrt{|g|}\varepsilon_{\mu\nu\rho},
\eeq
where $\eta_{\mu\nu\rho}$ and $\varepsilon_{\mu\nu\rho}$ are the Levi-Civita tensor and symbol, respectively,  and $\Lambda=-{1\over\ell^2}$. It is worth noting that the metric \eqref{WBH} becomes locally AdS$_3$ at $\mu={3\over \ell}$ (it is the BTZ black hole line element \cite{Banados:1992wn, Banados:1992gq}, albeit in unusual coordinates).

The metric \eqref{WBH} has been extensively studied in the literature, and boundary conditions have been outlined in previous accounts on the topic, \cite{Compere:2008cv, Compere:2009zj, Blagojevic:2009ek, Henneaux:2010fy, Henneaux:2011hv}. It belongs to a phase space whose asymptotic symmetry algebra is given by that of a WCFT in \emph{canonical} ensemble
\beqn \label{CSA}
 \left[L_{n}, L_{m}\right] &=& (n-m) L_{n+m}+\frac{c}{12} n(n-1)(n+1) \delta_{n+m},\\\left[L_{n}, P_{m}\right] &=&-m P_{m+n},  \\\left[P_{n}, P_{m}\right] &=&\frac{k}{2} n \delta_{n+m} , \label{CSAf}
\eeqn
where the central extensions are given by \cite{Anninos:2008fx, Compere:2009zj, Blagojevic:2009ek, Henneaux:2011hv}\footnote{In comparing with these references, we have different factors of $\ell$ because our time coordinate is dimensionless.}
\beqn \label{CentralExt}
 c =  \frac{15 \ell^2 \mu ^2+81}{G \ell^2 \mu ^3+27 G \mu },\qquad 
 k = -\frac{\ell^2 \mu ^2+27}{18 G \mu }.
\eeqn

For black hole solutions, the zero modes of the charges are given by
\beqn
L_0&=&Q_{\pa_\psi}= -{j\over 6G\mu^2\ell^2} \left(27+5 \ell^2 \mu ^2\right)+\frac{9 m^2}{2 G \mu },\\
P_0&=&Q_{\pa_T}= {\ell m\over G}.
\eeqn
Requiring the absence of naked singularities we find,  in terms of the parameters $m$ and $j$, that 
\beq\label{mc}
m^2 \geq \frac{j}{\ell^2 \mu }+\frac{j \mu }{27}=-{2 G k j\over 3 \ell^2}.
\eeq
This can be translated in terms of the zero mode charges as
\beq\label{Lc}
L_0 \geq{P_0^2\over k}.
\eeq

For these boundary conditions, we get a WCFT in canonical ensemble. The warped Cardy formula \cite{Detournay:2012pc}
 is
\beqn\label{Sent}
S =-{4\pi i P_0 P_0^{vac}\over k}+4\pi \sqrt{-\Big(L_0^{vac}-{(P_0^{vac})^2\over k}\Big)\Big(L_0-{P_0^{2}\over k}\Big)}.
\eeqn
The vacuum values of the charges were previously identified as
\beqn 
\quad L_0^{vac} = -\frac{c}{24} + \frac{(P_0^{vac})^2}{k},  \qquad
(P_0^{vac})^2 = - \frac{\ell^2}{36}.
\eeqn
Notice that  we have $c>0$ and thus
\beqn
L_0^{vac}-{(P_0^{vac})^2\over k}=-\frac{c}{24} < 0.
\eeqn
Consequently, the first condition for the well-definiteness of \eqref{Sent} is
\beq\label{Lcc}
L_0-{P_0^{2}\over k}> 0,
\eeq
which is automatically satisfied thanks to \eqref{Lc}. The result \eqref{Sent} was shown to match the Bekenstein-Hawking entropy of the corresponding dual black hole metrics in TMG. However, another condition on the spectrum necessary to derive \eqref{Sent} is that $L_0$ is bounded from below, see \cite{Detournay:2012pc}. Given \eqref{Lc}, and the fact that $j$ can attain any value, we are in a setup where $L_0$ is unbounded, which means that we are not in the regime of validity of the warped Cardy formula. This makes the use of the canonical ensemble in holographic instances questionable. 

In \cite{Detournay:2012pc}, WAdS$_3$ black holes were reinterpreted as a deformation of the BTZ metrics -- of the form 
\eqref{WBTZ}. The resulting spaces, already mentioned in the introduction and dubbed ``Warped BTZ" (WBTZ) are related to \eqref{WBH} through a charge-dependent change of coordinates. This change of coordinates induces a boundary change of coordinates between the vacuum metric on the plane (the warped version of the Poincare metric) and the black holes of Rindler type \cite{Detournay:2012pc, Detournay:2015ysa}, allowing an interpretation of the black holes as being dual to a thermal density matrix in the spirit of \cite{Maldacena:1998bw}. This change of coordinates has the effect of redefining the zero mode charges and suggests that the symmetry algebra differs from that of a WCFT in canonical ensemble. From the bulk gravitational field space viewpoint, this field-dependent change of coordinates leads to the same metric falloffs at the boundary but different varying quantities, that is, different boundary conditions.\footnote{An equivalent way to formulate the problem is to say that we are looking for the field-dependent behavior of the (ADM) lapse and shift for which the charges are integrable.} In particular, the level becomes charge-dependent (see also \cite{Apolo:2018eky} for a discussion), and the new zero-mode charges are related to the former ones by a quadratic redefinition (see Sect.4, eq.(55) of \cite{Detournay:2012pc}). A WCFT partition function involving these new charges is therefore said to be in \emph{quadratic ensemble}. In the latter, the conditions on the spectrum to ensure the validity of the warped Cardy formula are satisfied, as we review below. Nonetheless, intrinsic boundary conditions for the quadratic ensemble, containing these black hole solutions, have not been derived so far. As stated in the introduction, the main goal of this paper is to derive the aforementioned boundary conditions, such that the asymptotic symmetry algebra is that of a WCFT in quadratic ensemble and the spectrum is in the regime of validity of the warped Cardy formula, as just explained.

The change of coordinates aforementioned brings \eqref{WBH}  to the so-called WBTZ metric. Calling the new coordinates $(t,r,\phi)$, it is explicitly given by
\begin{eqnarray}
T(t,r,\phi)&=&\frac{2 \sqrt{2G\left(1-2 H^2\right) \left(2
   H^2+3\right) (L M-J)}}{\sqrt{3 L}L} \ t,\label{COC1}\\
\rho(t,r,\phi) &=&\frac{\sqrt{3 L} L}{4 \sqrt{ 2 G \left(2 H^2+3\right) (L
   M-J)}} r^2-\frac{\sqrt{3G L} J L^2}{\sqrt{2\left(2 H^2+3\right)
   (L M-J)}}, \\
\psi(t,r,\phi)&=&\frac{\sqrt{2 H^2+3} (L \phi +t)}{\sqrt{3} L^2}.\label{COC3}
\end{eqnarray}
The parameters $M, J, H,$ and $L$ just introduced are related to the WAdS$_3$ parameters as
\beqn
\ell =\frac{ \sqrt{3} L}{\sqrt{3+2H^2}},\quad
\mu= \frac{3 \sqrt{1-2 H^2}}{L},\quad
m=\sqrt{2} \sqrt{\frac{G L (L M-J)}{3(2 H^2+3)}},\quad
j=-\frac{3 G \sqrt{1-2 H^2} J L^2}{2 H^2+3}.
\eeqn
 The change of coordinates (\ref{COC1}-\ref{COC3}) is a non-local/charge dependent transformation. Indeed, defining $x^\pm={t\over L}\pm\phi$, we observe in particular that
\begin{eqnarray}
T=\frac{2 \sqrt{2} \sqrt{G} \sqrt{\sqrt{1-2 H^2} \left(2 H^2+3\right)}
   }{L^{3/2}} \sqrt{\tilde{{{P}}}_0} \ t,
\end{eqnarray}
where we introduced
\beq\label{PT0}
Q_{\pa_-}\equiv \tilde{{P}}_0=\frac{1}{3} \sqrt{1-2 H^2} (L M-J).
\eeq
The time coordinate is therefore rescaled by the charge. As we will shortly see, this affects the asymptotic symmetry algebra in a non-trivial way.

The resulting metric after the change of coordinates (\ref{COC1}-\ref{COC3}) is
\beqn \label{WBTZ2}
\D s^2_{WBTZ} &=& \frac{L^2 r^2}{16 G^2 J^2 L^2-8 G L^2 M r^2+r^4} \D r^2-2(H^2-{1\over 2})(4GL^2M-r^2)\D x^+\D x^-\\
&&-{8G^2L^2(J^2+(2H^2-1)L^2M^2)-8GH^2L^2Mr^2+H^2r^4\over 4GL(LM-J)}\D x^{+ 2}-2GL(2H^2-1)(LM-J)\D x^{-2}.\nonumber
\eeqn
This is the so-called WBTZ line element, which as advertised earlier can be obtained as a 
 deformation of the classic BTZ black hole spacetime \cite{Banados:1992wn, Banados:1992gq} of the form
\begin{equation} \label{WBTZ}
  \D s^2_{WBTZ} = \D s^2_{BTZ} - 2 H^2 \xi \otimes \xi    ,
\end{equation}
with
\beqn
\D s^2_{BTZ} = \Big( -{r^2\over L^2}+8MG\Big) \D t^2+{r^2L^2 \D r^2\over r^4+16 L^2 J^2G^2-8MGL^2 r^2}+8J G \D t\D \phi+r^2\D \phi^2,
\eeqn
the BTZ metric solving Einstein equations with $\Lambda=-{1\over L^2}$. The deformation $\xi$ is the one form dual under $\D s^2_{BTZ}$ to the vector field
\beq
\un \xi={1\over\sqrt{8 G L (LM-J)}}(-L \pa_t+\pa_\phi) = -{1\over\sqrt{2 G L (LM-J)}} \pa_- ,
\eeq
where the prefactor is obtained requiring the norm of this vector being $1$.

The metric \eqref{WBTZ} solves the TMG equations of motion \eqref{TMGeom} for\footnote{The signature here is $(-1,1,1)$ and we choose $\varepsilon^{tr\phi} = {1\over \sqrt{-g}}$. If one changes the convention on $\varepsilon$, i.e., changes the order of the coordinates, then one has to change the sign of $\mu$ accordingly.}
\beq
\Lambda =-{1\over 27 L^2}\big(36- \mu^2 L^2\big),\quad 
H^2 = {9-\mu^2 L^2\over 18}, \quad
\mu > 0.
\eeq

The metric has two globally defined Killing vectors $\pa_-$ and $\pa_+$ with corresponding conserved charges given by \eqref{PT0} and 
\begin{equation}\label{LT0}
  Q_{\pa_+}\equiv \tilde{{L}}_0= {2(1-H^2)\over 3\sqrt{1-2H^2}}{(LM+J)}.
\end{equation}

The family of solutions \eqref{WBTZ2} was conjectured in \cite{Detournay:2012pc} to be dual to thermal states in a WCFT in quadratic ensemble, that is a $2$-dimensional field theory with symmetry algebra
\beqn \label{QSA}
 \left[\tilde{L}_{n}, \tilde{L}_{m}\right] &=& (n-m) \tilde{L}_{n+m}+\frac{c}{12}\left(n^{3}-n\right) \delta_{n+m}, \\ \left[\tilde{L}_{n}, \tilde{P}_{m}\right] &=&-m \tilde{P}_{m+n},
 \\\left[\tilde{P}_{n}, \tilde{P}_{m}\right] &=&- 2 n \tilde{P}_{0} \delta_{m+n}. \label{QSAf}
\eeqn

The asymptotic degeneracy of states of field theories with algebras  (\ref{QSA}-\ref{QSAf}) has been determined in \cite{Detournay:2012pc} in a Cardy-like regime assuming certain conditions on the spectrum
\beqn \label{entropyQ}
S = 4 \pi \sqrt{-\tilde{P}_{0}^{v a c} \tilde{P}_{0}}+4 \pi \sqrt{-\tilde{L}_{0}^{v a c} \tilde{L}_{0}}.
\eeqn
Here, $\tilde{P}_{0}^{v a c}$ and $\tilde{L}_{0}^{v a c}$ are the zero modes of the vacuum of the theory in the quadratic ensemble. From the change of coordinates, the new zero modes are
\beqn
\tilde{L}_0 ={L}_0-{{P}_0^2\over k} , \qquad 
\tilde{ P}_0 =-{{P}_0^2\over k}.\label{Red0}
\eeqn
The condition on the spectrum is that $\tilde{L}_0$ must be bounded from below which, thanks to \eqref{Lcc}, is now satisfied in this ensemble, because $\tilde{L}_0\geq 0$. Hence, we are in the domain of validity, as claimed above. 

The goal of this manuscript is to determine a phase space in TMG with metrics falling off at infinity like the WBTZ metrics, which exhibit a different fall-off from BTZ due to the warping. Indeed, comparing \eqref{WBTZ2} with the unwarped BTZ metric,
\beq
\D s^2_{BTZ} = \frac{L^2 r^2}{16 G^2 J^2 L^2-8 G L^2 M r^2+r^4} \D r^2+(4GL^2M-r^2)\D x^+\D x^-+2GL(LM+J)\D x^{+ 2}+2GL(LM-J)\D x^{-2},
\eeq
we see that the warping is over-leading in the $\D x^+$ component. Eventually, this means that we are looking for relaxed boundary conditions, in order to capture WBTZ metrics, of the form 
\begin{equation}
g_{++}=O(r^4)  ,\quad g_{+-}=O(r^2) , \quad g_{--}=O(1), \quad g_{rr}={L^2\over r^2}+O(r^{-4}).
\end{equation}
and show that its symmetries are given by (\ref{QSA}-\ref{QSAf}).

\section{New boundary conditions}\label{s3}

Guided by the motivations of the previous section, we derive here new boundary conditions encompassing WBTZ black holes. We first outline the boundary conditions, then show how the solution space is restricted to give WBTZ black holes, and finally study two limits in which we retrieve familiar line elements.

\subsection{Setup}

A solution of \eqref{TMGeom} is given, trading $\Lambda$ for $L$ as
\beq
\Lambda=-{1\over 27 L^2}\big(36- \mu^2 L^2\big),\label{Lambda}
\eeq
and using the Fefferman-Graham (FG) gauge, with coordinates $x^a=(x^+,x^-)$ such that $x^\pm={t\over L}\pm \phi$,\footnote{One could have chosen any other length $\ell$ to normalize time. As long as we pick a quantity with dimension length it is the same. Note that $L$ becomes $\ell$, and so becomes the usual definition of $x^\pm$, only at $\mu L=\pm 3$. The conventions on $x^\pm$ are here the same as in \cite{Ciambelli:2020shy} (see [109] there).} 
\beq\label{AnG}
\D s^2=g_{\mu\nu}\D x^\mu \D x^\nu={L^2\over r^2}\D r^2+g_{ab}(r,x)\D x^a \D x^b,
\eeq
by the following metric components
\beqn
g_{++} &=& r^4 j_{++}+r^2 h(x^+)+f_{++}(x^+)+\frac{h(x^+) \left(\mu ^2 L^2-9\right) \left(4 j_{++} \mu ^2 L^2
   f_{++}(x^+)-h(x^+)^2 \left(\mu ^2 L^2-9\right)\right)}{8 r^2 j_{++}^2 \mu ^2
   L^2 \left(\mu ^2 L^2+9\right)}\nonumber\\
   &&+\frac{\left(\mu ^2 L^2-9\right)^2 \left(4 j_{++} \mu ^2 L^2 f_{++}(x^+)+h(x^+)^2
   \left(9-\mu ^2 L^2\right)\right)^2}{64 r^4 j_{++}^3 \mu ^4 L^4 \left(\mu ^2 L^2+9\right)^2} \label{gpp}\\ g_{+-}&=&-\frac{1}{18} \mu ^2 L^2 r^2-\frac{h(x^+)
   \left(\mu ^2 L^2-9\right)}{36 j_{++}}-\frac{\left(\mu ^2 L^2-9\right) \left(4 j_{++} \mu ^2 L^2 f_{++}(x^+)+h(x^+)^2
   \left(9-\mu ^2 L^2\right)\right)}{144 j_{++}^2 r^2 \left(\mu ^2 L^2+9\right)}\label{gpm}\\
   g_{--}&=&\frac{\mu ^2 L^2 \left(\mu ^2 L^2-9\right)}{324 j_{++}} \label{gmm}.
\eeqn
This metric solves \eqref{TMGeom}, with the convention $\varepsilon_{r + -}=-1$, which is consistent with $\varepsilon^{tr\phi}=1$. Note that the first term in \eqref{gpm} is the would-be boundary metric, which is thus flat. One could generalize this ansatz including a Weyl factor, and reproduce in this context what has been done in \cite{Alessio:2020ioh}. We will not do this enhancement here, leaving it for future work, and this will consequently freeze the dilatation symmetry in the asymptotic vectors below.

The parameter $L$ is from now on assumed to be positive, $L>0$, which implies that we have a solution of the equations of motion only in the range $\mu >0$.
The solution space is characterized by three quantities: a constant $j_{++}$ and two chiral functions $h(x^+)$ and $f_{++}(x^+)$.
Various comments are in order here. The first  is that the parameter $L$ becomes the usual AdS radius $\ell$ only for $\mu L= \pm{3}$. At these points indeed $\Lambda=-{1\over L^2}$. These values of $\mu$ are critical, and we will study shortly what happens there. In general, the $L$ appearing in the FG gauge is not the usual one, in contrast with what happens e.g. in \cite{Henneaux:2010fy, Henneaux:2011hv}. The second remark is that the line element has been directly defined on-shell, whereas it would have been more rigorous to find first appropriate boundary conditions off-shell. It can be viewed as a counterpart of the Ba\~nados metrics for AdS$_3$. 

It is important to note that the ansatz we made to solve the TMG equations of motion involved only integer powers of $r$. This in turn leads to a family of metrics that is locally equivalent to the Warped AdS$_3$ black holes, as can be checked by inspecting curvature invariants built out of the Ricci and Cotton tensors. Therefore, our analysis does not include the massive graviton around global WAdS$_3$. Including more general, non-diffeomorphic, solutions capturing the massive mode would require to introduce $mu$-dependent fall-offs, in the spirit of \cite{Henneaux:2009pw, Henneaux:2010fy}  for AdS$_3$ in TMG, and of \cite{Henneaux:2011hv} for WAdS$_3$. We leave this for future work.

Our boundary conditions  are interesting because they allow to navigate through the various space-like, time-like, and null  warped solutions. The WBTZ black holes in Fefferman-Graham gauge are given by
\beqn
\D s^2 &=&\frac{\D r^2 L^2}{r^2}+\frac{2}{9} \D x^{-2} G \mu ^2 L^3 (L M-J)-\frac{\D x^-\D x^+  \mu
   ^2 L^2 \left(4 G^2 L^2 (L^2 M^2-J^2)+r^4\right)}{9 r^2}\\
   &&+\frac{\D x^{+2} \left(\mu ^2 L^2 \left(4
   G^2 L^2  (L^2 M^2-J^2)+r^4\right)^2-9 \left(r^4-4 G^2 L^2  (L^2 M^2-J^2)\right)^2\right)}{72 G L r^4
   (L M-J)}.
   \label{WBTZFG} 
\eeqn
These solutions are included in our boundary conditions (\ref{gpp}-\ref{gmm}) for 
\beqn
j_{++} &=& \frac{\mu ^2 L^2-9}{72 G L (L M-J)}, \label{WBTZjpp} \\
h&=& 0,\label{WBTZh} \\
f_{++} &=& \frac{1}{9} G L \left(\mu ^2 L^2+9\right) (J+L M),\label{WBTZfpp}\\
\Delta&=&{\mu^2L^2-9 \over 144 G L j_{++}}= {(LM-J)\over 2}.\label{Delta}
\eeqn
Here we introduced also the quantity $\Delta$, which plays an important role in what follows.

Note that, given \eqref{Lambda}, if we want to keep the cosmological constant negative we have to assume $-6<\mu L<6$, which is further restricted to $0<\mu L<6$, since we assumed both $\mu$ and $L$ positive. The scalar curvature of \eqref{AnG} is given by
\beq
R={2(\mu L+6)(\mu L-6)\over 9L^2}=6\Lambda
\eeq
Consequently, it is negative as long as $\Lambda$ is.   Although extensions to asymptotically flat and WdS spacetimes are part of our agenda (see e.g. \cite{Anninos:2009jt}), from now on we will focus on these values of the Chern-Simons coupling.

\subsection{Limits}

Our new line element has two interesting limits. The first one is the null warped (NW) limit obtained setting $\mu L= 3$ while keeping $j_{++}$ arbitrary. The cosmological constant becomes $\Lambda=-{1\over L^2}=-{1\over \ell^2}$ and the line element collapses to
\beqn \label{NWL}
\D s^2={\ell^2\over r^2}\D r^2+(r^4 j_{++}+r^2 h(x^+)+f_{++}(x^+))(\D x^+)^2-r^2\D x^+\D x^-.
\eeqn
This metric is not an Einstein space, for the Cotton tensor has a non-vanishing component $C_{++}={12 j_{++} r^4\over \ell^3}$. The line element \eqref{NWL} falls inside a bigger class of boundary conditions, where the constant $j_{++}$ is allowed to be chiral $j_{++}(x^+)$ \cite{Anninos:2010pm}. See Appendix \ref{AppA} for further details.

The second limit is the CSS limit.  This limit involves both $\mu$ and $j_{++}$. It is performed sending $j_{++}$ to zero and $\mu L\to 3$ keeping their ratio constant, i.e. keeping $g_{--}$ finite in the limit. The quantity $\Delta={\mu^2L^2-9 \over 144 G L j_{++}}$ introduced in \eqref{Delta} is particularly relevant for this limit. Note that its sign depends on the sign of $j_{++}$. Using $\Lambda\to -{1\over L^2}=-{1\over \ell^2}$, we obtain
\beqn \label{CSS}
\D s^2&=&{\ell^2\over r^2}\D r^2+\Big(r^2 h(x^+)+f_{++}(x^+)-\frac{4 \Delta G \ell  h(x^+) (4 \Delta G \ell  h(x^+)^2- f_{++}(x^+))}{r^2}\Big)(\D x^+)^2+4\Delta G \ell(dx^-)^2\nonumber\\
&&-\Big(r^2+8\Delta G \ell h(x^+)-\frac{4\Delta G \ell  (4 \Delta G \ell h(x^+)^2- f_{++}(x^+))}{r^2}\Big)\D x^+\D x^-.
\eeqn
We call this the CSS limit because the metric falls into CSS boundary conditions \cite{Compere:2013bya}, embedded in TMG as in \cite{Ciambelli:2020shy}. This metric has identically vanishing Cotton tensor, so it is a solution of Einstein equations. The exact matching between \eqref{CSS} and the CSS line element (eq. (1) in \cite{Ciambelli:2020shy}) is made via the identification
\beqn \label{CSSDict}
h(x^+)= \pa_{+}P(x^+),\qquad
f_{++}(x^+)= 4 G \ell (\bar{L}(x^+)+\Delta  (\pa_+ P(x^+))^2).
\eeqn
We have therefore introduced new boundary conditions that encompass previous results. Furthermore, they are useful because the asymptotic symmetry algebra turns out to be that of a WCFT in quadratic ensemble, as we now demonstrate.

We may also observe that the metric \eqref{WBTZFG} becomes ill-defined when $J = M L$. However, as we will see in the forthcoming sections and can already be seen on \eqref{PT0} and \eqref{LT0}, the charges themselves are well-defined at these values of the parameters, indicating a mere coordinate singularity. The metric can be made regular and non-degenerate by a rescaling  $r \rightarrow  (-J + L M)^{1/4} r$ and $x^-  \rightarrow (-J + L M)^{-1/2} x^-$, and still belonging to our family of metrics.

\section{Symmetries and charges}\label{s4}

We begin this section with the analysis of residual symmetries and the most general variation of solution space. We then show that surface charges are in general not integrable, and discuss two possible consistent sets of boundary conditions: a restriction of the solution space, or a field-dependent redefinition of the generators. In both cases, we study the charge algebra and consider WBTZ black holes.

\subsection{Residual symmetries and variation of solution space}

We denote by $\chi$ the solution space, $\chi=\{j_{++},h(x^+),f_{++}(x^+)\}$. We impose to preserve the FG gauge
\beqn
\Lie_{\un\xi}g_{rr}=0=\Lie_{\un\xi}g_{ra},
\eeqn
where $\Lie$ denotes the Lie derivative.
As reviewed in \cite{Ciambelli:2020eba, Ciambelli:2020ftk}, the most general solution is simply
\beqn
\xi^r = r\eta(x^+,x^-),\qquad
\xi^b = \xi^b_0(x^+,x^-)-L^2 \pa_a\eta(x^+,x^-)\int_\infty^r \D r^\prime {g^{ab}(x^+,x^-)\over r^\prime},
\eeqn
and depends on three arbitrary functions $\eta(x^+,x^-)$ and $\xi^b_0(x^+,x^-)$. Given the discussion above, it is natural to furthermore impose that residual symmetries leave the metric chiral:
\beqn
\xi^r = r\eta(x^+),\qquad
\xi^b = \xi^b_0(x^+)-L^2 \pa_a\eta(x^+)\int_\infty^r \D r^\prime {g^{ab}(x^+)\over r^\prime}.
\eeqn
To avoid confusion, and because everything is chiral, we call $\xi^+_0(x^+)=\epsilon$ and $\xi^-_0(x^+)=\sigma$ in the following. Given our line element, we obtain :
\beqn\label{AKV}
\un\xi&=&\xi^r\pa_r+\xi^+\pa_++\xi^-\pa_-\\
&=& r\eta \pa_r+\Big(\epsilon-\frac{2 j_{++} \mu ^2 L^4 (\mu ^4 L^4-81) \eta^\prime}{9 (-4 j_{++} \mu^2 L^2 f_{++} (\mu ^2 L^2-9)+h^2 (\mu ^2 L^2-9)^2+8
j_{++}^2 \mu ^2 L^2 r^4 (\mu ^2 L^2+9))}\Big)\pa_+\nonumber\\
&& +\Big(\sigma-\frac{2 j_{++} L^2 (\mu ^2 L^2+9) \eta' (h (\mu ^2 L^2-9)+4j_{++} \mu ^2 L^2 r^2)}{-4 j_{++} \mu ^2 L^2 f_{++}(\mu ^2 L^2-9)+h^2 (\mu ^2 L^2-9)^2+8 j_{++}^2 \mu ^2 L^2 r^4(\mu ^2 L^2+9)}\Big)\pa_-.
\eeqn
We remark that, since the line element has a finite expansion in the coordinate $r$, these vector fields are written in closed form in $r$.

We now proceed to compute the variation of the solution under residual symmetries $\delta_{\un\xi}\chi$ using as usual $\delta_{\un\xi}g_{\mu\nu}=\Lie_{\un\xi}g_{\mu\nu}$, expanded in powers of $r$. We first focus on the leading term in the $++$ component:
\beq
\delta_{\un\xi}j_{++}=2 j_{++}(2 \eta(x^+)+\epsilon^\prime(x^+)).
\eeq
We then require $j_{++}$ to remain constant, which implies
\beq
\eta=-{1\over 2} \epsilon^\prime+\eta_0,
\eeq
with $\eta_0$ a constant transforming $j_{++}$:
\beq
\delta_{\un\xi}j_{++}=4j_{++} \eta_0.
\eeq
Although this is admissible for $j_{++}$, if we compute the $+-$ component we obtain
\beq
\Lie_{\un\xi}g_{+-}=-r^2{\mu^2 L^2 \eta_0 \over 9}+O(r^0),
\eeq
which is not allowed, and thus we must impose $\eta_0=0$. It would be interesting to enhance this construction including an arbitrary boundary conformal factor as done in \cite{Alessio:2020ioh} for Einstein gravity, allowing us to make contact with the recent work \cite{Chaturvedi:2020jyy}. Keeping $\eta_0=0$ means that $j_{++}$ is fixed along the residual orbits. We proceed and find the full residual variation of $\chi$, that we summarize here:
\beqn
\delta_{\un\xi}j_{++}&=& 0,\\
\delta_{\un\xi}h&=& \epsilon h^\prime+h \epsilon^\prime-\frac{1}{9} \mu ^2 L^2
   \sigma^\prime \label{dh},\\
\delta_{\un\xi}f_{++}&=& \epsilon f_{++}^\prime+2f_{++} \epsilon^\prime-{(\mu ^2 L^2-9)h \sigma^{\prime}\over 18j_{++}}-{(\mu ^2 L^2+9)L^2 \epsilon^{\prime\prime\prime}\over 36}.
\eeqn
The second expression suggests that $h$ is a $u(1)$ current with level related to the last term. The last expression indicates that $f_{++}$ is a Virasoro current, where the last term is the one related to the central extension.

The most general residual symmetries are thus generated by the on-shell vectors:
\beqn
\un\xi&=&\xi^r\pa_r+\xi^+\pa_++\xi^-\pa_-,\\
&=& \nonumber-{r\over 2}\epsilon^{\prime} \pa_r+\Big(\epsilon+\frac{ j_{++} \mu ^2 L^4 (\mu ^4 L^4-81) \epsilon^{\prime\prime}}{9 (-4 j_{++} \mu^2 L^2 f_{++} (\mu ^2 L^2-9)+h^2 (\mu ^2 L^2-9)^2+8
j_{++}^2 \mu ^2 L^2 r^4 (\mu ^2 L^2+9))}\Big)\pa_+\\
&& +\Big(\sigma+\frac{ j_{++} L^2 (\mu ^2 L^2+9) \epsilon^{\prime\prime} (h (\mu ^2 L^2-9)+4j_{++} \mu ^2 L^2 r^2)}{-4 j_{++} \mu ^2 L^2 f_{++}(\mu ^2 L^2-9)+h^2 (\mu ^2 L^2-9)^2+8 j_{++}^2 \mu ^2 L^2 r^4(\mu ^2 L^2+9)}\Big)\pa_-,
\eeqn
which depend on two arbitrary chiral functions $\epsilon(x^+)$ and $\sigma(x^+)$. For the asymptotic symmetry algebra, $\sigma$ generates an abelian algebra
\beq
[\un\sigma^1,\un\sigma^2]=0, \quad \text{with} \quad \un\sigma^i=\sigma_i(x^+)\pa_-,
\eeq
while $\epsilon$ generates the usual Witt algebra, under the modified Lie brackets, see \cite{Barnich:2010eb}, due to the field dependence of the vectors. The total asymptotic symmetry algebra is therefore a semi-direct sum of a Witt and a $u(1)$ algebra.

The notation we will adopt is
\beqn
\un\sigma&=&\sigma(x^+)\pa_-,\\
\un\epsilon&=&-{r\over 2}\epsilon^{\prime} \pa_r+\Big(\epsilon+\frac{ j_{++} \mu ^2 L^4 (\mu ^4 L^4-81) \epsilon^{\prime\prime}}{9 (-4 j_{++} \mu^2 L^2 f_{++} (\mu ^2 L^2-9)+h^2 (\mu ^2 L^2-9)^2+8
j_{++}^2 \mu ^2 L^2 r^4 (\mu ^2 L^2+9))}\Big)\pa_+ \nonumber\\
&& +\frac{ j_{++} L^2 (\mu ^2 L^2+9) \epsilon^{\prime\prime} (h (\mu ^2 L^2-9)+4j_{++} \mu ^2 L^2 r^2)}{-4 j_{++} \mu ^2 L^2 f_{++}(\mu ^2 L^2-9)+h^2 (\mu ^2 L^2-9)^2+8 j_{++}^2 \mu ^2 L^2 r^4(\mu ^2 L^2+9)}\pa_-.
\eeqn
At this stage, we have found the variation of solution space and the residual symmetries, so we have all the ingredients to compute surface charges.

\subsection{Surface charges and algebra}

We start with the $u(1)$ sector and compute TMG surfaces charges. We obtain that the Iyer-Wald \cite{Wald:1993nt, Iyer:1994ys, Bouchareb:2007yx, Compere:2008cv} charges read\footnote{In this paper, charges are computed using the package \cite{code}.}
\beq
\cancel\delta Q_{\un\sigma}[g,h]=-\frac{\mu ^2 L^2-9}{1296 \pi  G \mu  L^2 j_{++}^2}\int_0^{2\pi} \D \phi \ \sigma(x^+)\Big(\delta j_{++} (\mu ^2
   L^2+9 h(x^+))-18 j_{++} \delta h(x^+)\Big) \label{U1Charge}.
\eeq
Here we have computed the surface charges first at $(r,x^-)$ fixed, then at $(r,x^+)$ fixed,  added them and sent $r\to \infty$. These charges are finite, conserved but integrable only if $\delta j_{++}=0$. However, as we will see one can find combination of residual vectors such that these charges become integrable even when $\delta j_{++}\neq 0$. Although there they are integrable, we observe some similarities with \cite{Ciambelli:2020shy} and \cite{Alessio:2020ioh}, where one needs to find specific orbits of the residual symmetry vectors to obtain a direct sum algebra. Indeed, in cases where the charges are not integrable, such redefinition makes them so, as discussed in \cite{Adami:2020ugu, Ruzziconi:2020wrb, Adami:2021sko, Adami:2021nnf}.

Before proceeding, we report here also the Virasoro charges
\beqn
\cancel\delta Q_{\un\epsilon}[g,h]&=&\nonumber {1\over 144 \pi G L^4\mu^3 j_{++}^2}\int^{2\pi}_0\D \phi\Big(27(\mu^2L^2 -9)h^2\epsilon\delta j_{++}+(\mu^2L^2-9)h\epsilon(\mu^2L^2\delta j_{++}-54 j_{++}\delta h)\\
&&+\mu^2L^2j_{++}\Big(L^2(\mu^2L^2-9)\delta j_{++}\epsilon^{\prime\prime}+72j_{++}\epsilon\delta f_{++}\Big)\Big) \label{VirasoroCharge}.
\eeqn
where we used
\beqn
\un\epsilon&=&\nonumber-{r\over 2}\epsilon^\prime \pa_r+\Big(\epsilon+\frac{ j_{++} \mu ^2 L^4 (\mu ^4 L^4-81) \epsilon^{\prime\prime}}{9 (-4 j_{++} \mu^2 L^2 f_{++} (\mu ^2 L^2-9)+h^2 (\mu ^2 L^2-9)^2+8
j_{++}^2 \mu ^2 L^2 r^4 (\mu ^2 L^2+9))}\Big)\pa_+\\
&& +\Big(\frac{ j_{++} L^2 (\mu ^2 L^2+9) \epsilon^{\prime\prime} (h (\mu ^2 L^2-9)+4j_{++} \mu ^2 L^2 r^2)}{-4 j_{++} \mu ^2 L^2 f_{++}(\mu ^2 L^2-9)+h^2 (\mu ^2 L^2-9)^2+8 j_{++}^2 \mu ^2 L^2 r^4(\mu ^2 L^2+9)}\Big)\pa_-.
\eeqn

We would then then to compute the charge algebra. To do so, we need to find a set of integrable charges. As we discussed, there exist two possibilities that do not further constraint the residual vectors
\begin{enumerate}
\item Set $\delta j_{++}=0$ on the solution space. 
\item Perform a field dependent redefinition of the generators of residual symmetries $\sigma$ and $\epsilon$.
\end{enumerate}
Both possibilities have interesting consequences, so we study them in detail in the following subsections.

\subsubsection{The case $\delta j_{++}=0$ ($\Delta$ fixed)}

A simple situation is attained imposing $\delta j_{++}=0$, which does not further reduce the residual symmetries. The charges then read
\beqn\label{u1ch}
\cancel\delta Q_{\un\sigma}[g,h] &=& \frac{\mu ^2 L^2-9}{72 \pi  G \mu  L^2 j_{++}}\int_0^{2\pi} \D \phi \ \sigma(x^+)\delta h(x^+),\\\label{virch}
\cancel\delta Q_{\un\epsilon}[g,h]&=& {1\over 8 \pi G L^4\mu^3 j_{++}}\int^{2\pi}_0\D \phi \ \epsilon(x^+)\Big(4\mu^2L^2j_{++}\delta f_{++}-3 (\mu^2L^2-9)h\delta h\Big).
\eeqn
These can now be integrated up to an arbitrary background value (integration constant). We fix this by demanding that the zero mode of the charge $Q_{\un\sigma}[g]$ matches its zero mode when $j_{++}$ is allowed to vary, as also done in \cite{Compere:2013bya}. This also ensures that we reproduce the charges for the exact isometries of WBTZ black holes. We thus get
\beqn\label{qs}
Q_{\un\sigma}[g] &=& \frac{\mu ^2 L^2-9}{72 \pi  G \mu  L^2 j_{++}}\int_0^{2\pi} \D \phi \ \sigma( h +{\mu^2L^2\over 18}),\\
Q_{\un\epsilon}[g]&=&{1\over 16 \pi G L^4\mu^3 j_{++}}\int^{2\pi}_0\D \phi \ \epsilon\Big(8\mu^2L^2j_{++} f_{++}-3 (\mu^2L^2-9) h^2\Big) \label{quadqe} .
\eeqn

 We can therefore directly compute the charge algebra here. We start with the $u(1)$ sector and get
\beqn
\{ Q_{\un\sigma^1}[g], Q_{\un\sigma^2}[g] \}=\delta_{\un\sigma^2}Q_{\un\sigma^1}[g]=\frac{\mu ^2 L^2-9}{72 \pi  G \mu  L^2 j_{++}}\int_0^{2\pi} \D \phi \ \sigma_1 \delta_{\un
\sigma^2}h= -\frac{\mu(\mu ^2 L^2-9)}{648 \pi  G  j_{++}}\int_0^{2\pi} \D \phi \ \sigma_1 
\sigma_2^\prime\label{u1j0}.
\eeqn
Since $Q_{[\un\sigma^1,\un\sigma^2]_M}[g]=0$, \eqref{u1j0} is the central extension for the $u(1)$ sector. Using the mode decomposition representation $\sigma_1=e^{i m x^+}$ and $\sigma_2=e^{i n x^+}$, and calling $Q_{\un\sigma^1}[g]=\tilde{{P}}_m$ and $Q_{\un\sigma^2}[g]=\tilde{{P}}_n$:
\beqn
\tilde{{P}}_m=\frac{\mu ^2 L^2-9}{72 \pi  G \mu  L^2 j_{++}}\int_0^{2\pi} \D \phi \ e^{i m x^+}  (h+{\mu^2L^2\over 18}){=\frac{2\Delta}{\mu L \pi}\int_0^{2\pi} \D \phi \ e^{i m x^+}(h+{\mu^2L^2\over 18})},
\eeqn
we obtain\footnote{Conventions: $\int_0^{2\pi} \D \phi e^{i (m+n) \phi}=2\pi \delta_{m+n,0}$.}
\beq
i\{\tilde{{P}}_m,\tilde{{P}}_n\}= m {\tilde{k}\over 2} \delta_{m+n,0}, \qquad \tilde{k}=-\frac{\mu(\mu ^2 L^2-9)}{162   G  j_{++}}=-\frac{8\mu L}{9}\Delta.
\eeq
This is a centrally extended $u(1)$ algebra with central extension $\tilde{k}$ -- colloquially called Kac-Moody level. We want to make connection with the quadratic algebra \eqref{QSA} and that the zero modes of our metric represent WBTZ black holes. This requires that $h(x^+)$ doesn't have a zero mode (see \eqref{WBTZh}) which can be achieved by demanding that $h(x+)=\pa_+H(x^+)$ for some arbitrary periodic function $H(x^+)$. This is analogous to the discussion in \cite{Compere:2013bya}, where one of the arbitrary chiral function that appears in the metric is $\pa_+ P(x^+)$,  ensuring that the  zero modes of the CSS metric coincides with those of the BTZ black hole 
 (see \eqref{CSSDict})\footnote{Also, note that (2.4)-(2.6) of \cite{Apolo:2018eky} is correct only if $H(x^+)$ is periodic.}.
  Thus, the condition $h(x^+)=\pa_+H(x^+)$ with $ H(x^+) $ periodic implies that
\beq \label{QLevel}
 \tilde{k} = -4 \tilde{{P}}_0.
\eeq

For the Virasoro sector we have:
\beqn
\{ Q_{\un\epsilon^1}[g], Q_{\un\epsilon^2}[g] \} = \nonumber\delta_{\un\epsilon^2}Q_{\un\epsilon^1}[g]&=&{1\over 16 \pi G L^4\mu^3 j_{++}}\int^{2\pi}_0\D \phi \Big(\Big(\epsilon_1\epsilon^{\prime}_2-\epsilon_2\epsilon^{\prime}_1\Big)\Big(8\mu^2L^2j_{++} f_{++}-3 (\mu^2L^2-9) h^2\Big)\\
&&-{(\mu^2L^2+9)\over 72\pi G \mu}\int^{2\pi}_0\D \phi \epsilon_2^{\prime\prime\prime}\epsilon_1.
\eeqn
The first line in this expression is clearly $Q_{[\un\epsilon^1,\un\epsilon^2]_M}[g]$, where one should use the modified Lie bracket
\begin{align}
\big[\un\epsilon^1,\un\epsilon^2\big]_{M}:=\big[\un\epsilon^1,\un\epsilon^2\big]-\delta_{\un\epsilon^1}\un\epsilon^2+\delta_{\un\epsilon^2}\un\epsilon^1,\label{mod}
\end{align}
because the vectors are field dependent, while the second line is the central extension. Using the mode decomposition representation $\epsilon_1=e^{i m x^+}$ and $\epsilon_2=e^{i n x^+}$, and calling $Q_{\un\epsilon^1}[g]=\tilde{{L}}_m$ and $Q_{\un\epsilon^2}[g]=\tilde{{L}}_n$:
\beq
\tilde{{L}}_m={1\over 16 \pi G L^4\mu^3 j_{++}}\int^{2\pi}_0\D \phi \ e^{i m x^+}\Big(8\mu^2L^2j_{++} f_{++}-3 (\mu^2L^2-9) h^2\Big),
\eeq
we obtain
\beq
i\{\tilde{{L}}_m,\tilde{{L}}_n\}=(m-n) \tilde{{L}}_{m+n}+{c\over 12}m^3 \delta_{m+n,0}, \qquad c=\frac{\mu ^2 L^2+9}{3 G \mu}.
\eeq
Note that both central charges are sensitive to the sign of $\mu$. 

Finally the mixed sector, with the conventions established, reads
\beq
i \{\tilde{{L}}_m,\tilde{{P}}_n\}=-n \tilde{{P}}_{m+n},
\eeq
which is the expected semi-direct action of the Virasoro sector on the Abelian one.

To summarize, we have obtained the algebra
\beqn
i\{\tilde{{L}}_m,\tilde{{L}}_n\} &=& (m-n) \tilde{{L}}_{m+n}+{c\over 12}m^3 \delta_{m+n,0},\\
i \{\tilde{{L}}_m,\tilde{{P}}_n\} &=& -n \tilde{{P}}_{m+n},\\
i\{\tilde{{P}}_m,\tilde{{P}}_n\} &=& m {\tilde{k}\over 2} \delta_{m+n,0},
\eeqn
with central extensions
\beqn
c = \frac{\mu ^2 L^2+9}{3 G \mu},\qquad
\tilde{k} = -\frac{\mu(\mu ^2 L^2-9)}{162   G  j_{++}}=-\frac{8\mu L}{9}\Delta,
\eeqn
where we recall $\Delta={\mu^2L^2-9 \over 144 G L j_{++}}$.
This algebra is the one of the centrally extended group 
\beq
Vir\ltimes U(1).
\eeq
While $c$ is field independent, the $U(1)$ central extension $\tilde{k}$ depends on $j_{++}$. With \eqref{QLevel}, the algebra precisely becomes (\ref{QSA}-\ref{QSAf}), confirming the expectation that our phase space has the same symmetries as that of a WCFT in quadratic ensemble. This is why we used the tilde notation for the generators.

Let us study the Null Warped limit, reached simply taking $\mu L\to 3$. In this case the $u(1)$ level and charges vanish identically, and we are left with a Virasoro symmetry algebra with central extension 
\beq
c=\frac{2 \ell}{G} \quad \text{at} \quad \mu L \to 3,
\eeq
where again $L\to \ell$, the AdS radius, in this limit. The disappearance of the $u(1)$ is the reason why we typically think of null warped solutions as extremal. This is also suggested by the fact that the chiral function $h$ disappears from the residual vectors in \eqref{AKV} and the charges (\ref{u1ch}-\ref{virch}) in the limit, where therefore it is a pure gauge datum. 

As we have seen, the CSS limit is reached setting $\mu L\to  3$ and $j_{++}\to 0$ while keeping ${\mu^2L^2-9 \over 144 G L j_{++}}=\Delta$ constant. The charges read
\beqn
Q_{\un\sigma}[g] =  \frac{2\Delta}{3\pi} \int_0^{2\pi} \D \phi \ \sigma( h+{1\over 2}),\qquad
Q_{\un\epsilon}[g]  = {1\over 6 \pi G \ell}\int^{2\pi}_0\D \phi \ \epsilon\Big(f_{++}-6 \Delta \ell G h^2\Big),
\eeqn
while the central extensions become
\beq
\tilde{k}=- \frac{8\Delta}{3}, \qquad  c=\frac{2\ell}{G}.
\eeq
Thus, this limit coincides with our findings in \cite{Ciambelli:2020shy}, where, for generic $\mu$, the central extensions were found to be
\beqn
k_{\text{CSS}} = -\big( 1-{1\over\mu\ell}\big)4\Delta,\quad
c_{\text{CSS}} = \big( 1+{1\over\mu\ell}\big) \frac{3\ell}{2G}.
\eeqn
We remark that the CSS metrics, even when embedded in TMG, have vanishing Cotton, so the equations of motion are invariant under chiral parity transformations. On the other hand, our boundary conditions admit solutions with non-vanishing Cotton so they are sensitive to chirality. This ultimately constraints the sign of $\mu$ (here assumed positive) as we previously saw.

We now turn our attention to WBTZ black holes.
These are reached restricting the solution space to
\beqn
j_{++} = \frac{\mu ^2 L^2-9}{72 G L (L M-J)}, \quad
h= 0,\quad
f_{++} = \frac{1}{9} G L \left(\mu ^2 L^2+9\right) (J+L M),\quad
\Delta={\mu^2L^2-9 \over 144 G L j_{++}}= {(LM-J)\over 2}.
\eeqn
Therefore, their charges in the quadratic ensemble take the form
\beqn
\tilde{{P}}_m = \frac{(LM-J) \mu L}{9}\delta_{m,0} \label{tpo}, \qquad
\tilde{{L}}_m = (\mu^2L^2+9){(LM+J)\over 9  L\mu } \delta_{m,0}.
\eeqn 
with $M$ and $J$ the Einstein charges.
The TMG mass and angular momentum of these solutions are defined as
\beqn
\tilde{M} = Q_{\pa_t} = {1\over L}\Big(Q_{\pa_+}+Q_{\pa_-}\Big), \qquad
\tilde{J} = Q_{\pa_\phi}= Q_{\pa_+}-Q_{\pa_-},
\eeqn
and we also have
\beqn
Q_{\pa_-}= \tilde{{P}}_0,\qquad
Q_{\pa_+}= \tilde{{L}}_0.
\eeqn
Subsequently, we obtain the relationship between the TMG mass and angular momentum and the zero modes of the charges
\beqn
\tilde{M} =  {1\over L}\Big(\tilde{{L}}_0+\tilde{{P}}_0\Big) ={2\mu L M\over 9}+{LM+J\over \mu L^2}\label{tM},\qquad
\tilde{J} = \tilde{{L}}_0-\tilde{{P}}_0={2\mu L J\over 9}+{LM+J\over \mu L}.\label{tL}
\eeqn

For BTZ, it has been found in \cite{Ciambelli:2020shy} that\footnote{In comparing, we should change the sign of $J$  and send the charge for $\pa_\phi$ to the charge for $-\pa_\phi$.}
\beqn
Q_{\un\sigma}[g] =  \frac{1}{2\pi \mu \ell} \int_0^{2\pi} \D \phi \ \sigma (\mu \ell-1)\Delta,\qquad
Q_{\un\epsilon}[g]  = \frac{1}{2\pi \mu \ell} \int_0^{2\pi} \D \phi \ \epsilon (\mu \ell+1)\bar\Delta,
\eeqn
with $\Delta={1\over 2}(LM-J)$, $\bar\Delta={1\over 2}(LM+J)$, and $f_{++}=4GL\bar\Delta$. At $\mu\ell=3$, this consistently coincides with the results above. In particular, the mass and angular momentum \eqref{tM} in the CSS limit at $\mu L=3$  read
\beqn
\tilde{M}  = M+{J\over 3 \ell},\qquad 
\tilde{J} = J+{\ell M\over 3},
\eeqn
which coincide exactly with $(21)$ and $(22)$ of  \cite{Ciambelli:2020shy} for $\mu\ell=3$, if we consistently change the sign of $J$ and $\tilde J$, because there we used $\tilde J=Q_{-\pa_\phi}$.

To summarize, we showed in this subsection that the bulk solution space has a symmetry algebra identified with that of a WCFT in quadratic ensemble.

\subsubsection{Varying $\Delta$ and generators redefinition}

Another set of boundary conditions that makes our charges integrable is achieved performing a redefinition of the generators of residual symmetries following CSS
 \cite{Compere:2013bya}
\beqn
\un\sigma \to  \un{\bs}={3 \un\sigma \over  \mu L \sqrt{|\Delta|}} \quad \Rightarrow \quad \bs={3 \sigma \over  \mu L \sqrt{|\Delta|}}, \qquad
\un\epsilon \to  \un\be=\un\epsilon-\epsilon(x^+)\pa_- \quad \Rightarrow \quad \be={\epsilon}.
\eeqn
Here we took the absolute value of $\Delta$ because this constant can assume both positive and negative values. So the two new chiral vector fields are $\un\be$ and $\un
\bs$, and depend on two arbitrary chiral functions $\epsilon(x^+)$ and $\sigma(x^+)$.
We are allowed to do so because all the generators are functions of $x^+$, so the Fourier decomposition is the same. Furthermore, the redefinition of $\be$ uses the fact that we subtract to the old one the $u(1)$ generator, which eventually keeps the algebra invariant.

The $u(1)$ charges then become
\beqn
\cancel\delta Q_{\un\bs}[g,h] &=& -\frac{\mu ^2 L^2-9}{1296 \pi  G \mu  L^2 j_{++}^2}\int_0^{2\pi} \D \phi \ \bs(x^+)\Big(\delta j_{++} (\mu ^2
   L^2+9 h(x^+))-18 j_{++} \delta h(x^+)\Big)\\
   &=&-\frac{(\mu ^2 L^2-9) }{36 \pi  \sqrt{GL^5} \mu^2  |j_{++}|\sqrt{|j_{++}|}\sqrt{|\mu^2L^2-9|}} \int_0^{2\pi} \D \phi \ \sigma(x^+)\Big(\delta j_{++} (\mu ^2
   L^2+9 h(x^+))-18 j_{++} \delta h(x^+)\Big)\nonumber,
\eeqn
which are easily integrated in solution space. 
\beqn
Q_{\un\bs}[g]&=&sg(j_{++})\frac{(\mu ^2 L^2-9)}{18 \pi  \sqrt{GL^5} \mu^2\sqrt{|\mu^2L^2-9|}\sqrt{|j_{++}|}} \int_0^{2\pi} \D \phi \ \sigma(x^+) \Big(\mu^2 L^2+9h(x^+)\Big) \nonumber\\
&=&sg(\Delta)\frac{2\sqrt{|\Delta|}}{3 \pi L^2 \mu^2}\int_0^{2\pi} \D \phi \ \sigma(x^+) \Big(\mu^2 L^2+9h(x^+)\Big),
\eeqn
where we use that $sg(\Delta)=sg(j_{++})sg(\mu ^2 L^2-9)$.
Similarly, for the Virasoro sector the charges become
\beqn
\cancel\delta Q_{\un\be}[g,h]
&=&\nonumber {1\over 1296 \pi G L^4\mu^3 j_{++}^2}\int^{2\pi}_0\D \phi \ \epsilon(x^+)\Big((\mu^2L^2-9)\delta j_{++}(\mu^4 L^4+18 \mu^2L^2 h(x^+)+243 h(x^+)^2)\\&&+18 j_{++}\big(36\mu^2 L^2 j_{++}\delta f_{++}(x^+)-(\mu^2L^2-9)(\mu^2L^2 +27 h(x^+)) \delta h(x^+) \big) \Big),
\eeqn
where we have thrown away total derivatives. Also this can be integrated in solution space. Trading $j_{++}$ for $\Delta$, which organizes the expressions better, the result is
\beqn
  Q_{\un\be}[g]=\frac{1}{18 \pi  G \mu ^3 L^4}\int_0^{2\pi}\D \phi \ \epsilon(x^+) \Big(- 2GL\Delta(\mu ^4 L^4+18 \mu ^2 L^2
  h(x^+) +243h(x^+) ^2)+9 \mu ^2 L^2 f_{++}(x^+) \Big).
\eeqn

The charge algebra, calling $\un\bx^i=(\un\be^i,\un\bs^i)$ and given the new gauge orbits, is given by
\beq
\{ Q_{\un\bx^1}[g], Q_{\un\bx^2}[g] \}=\delta_{\un\bx^2}Q_{\un\bx^1}[g]=Q_{[\un\bx^1,\un\bx^2]_M}[g]+{\cal K}_{\un\bx^1,\un\bx^2},
\eeq
where ${\cal K}_{\un\bx^1,\un\bx^2}$ is the central extension. In this expression, since the generators are now field dependent, one should use the modified bracket \eqref{mod}. However, since $\delta_{\un\xi}j_{++}=0$, in this specific instance the modified bracket coincides with the ordinary Lie bracket. For the $u(1)$ sector we gather
\beq\label{u1mod}
\{ Q_{\un\bs^1}[g], Q_{\un\bs^2}[g] \}=\delta_{\un\bs^2}Q_{\un\bs^1}[g]=-sg(\Delta)\frac{2}{ \pi \mu L}\int_0^{2\pi} \D \phi \ \sigma_1
\sigma_2^{\prime}.
\eeq
Since $Q_{[\un\bs^1,\un\bs^2]_M}[g]=0$, \eqref{u1mod} is the central extension for the $\hat u(1)$ sector. Using the mode decomposition representation $\sigma_1=e^{i m x^+}$ and $\sigma_2=e^{i n x^+}$, and calling $Q_{\un\bs^1}[g]={{P}}_m$ and $Q_{\un\bs^2}[g]={{P}}_n$:\footnote{Note that the new vector fields are written with respect to  $\bs$, but the field independent quantity is $\sigma$, which can thus be expanded in modes.}
\beq
{{P}}_m=sg(\Delta)\frac{2\sqrt{|\Delta|}}{3 \pi L^2 \mu^2}\int_0^{2\pi} \D \phi \ e^{imx^+}\Big(\mu^2 L^2+9h(x^+)\Big),
\eeq
we obtain\footnote{At $\mu L=3$, this result is the analogue of $(13)$ of \cite{Ciambelli:2020shy} but for the canonical ensemble, namely, the level is equal to $\Big((1-{1\over \mu \ell}) \text{CSS}\Big)|_{\mu \ell\to 3}$, where CSS stands for the CSS level in Einstein gravity in canonical ensemble (as in App. B of \cite{Compere:2013bya}).}
\beq
i\{{{P}}_m,{{P}}_n\}= m {k\over 2} \delta_{m+n,0}, \qquad k=-sg(\Delta)\frac{8}{\mu  L}.
\eeq
This is a centrally extended $\hat u(1)$ algebra with central extension (level) $k$. As anticipated, this level coincides exactly with \eqref{CentralExt}, except for the $sg(\Delta)$ term, determining the sector 
of the theory. We remark that such extension depends on the sign of $j_{++}$, which is the warped equivalent of the result in \cite{Compere:2013bya} for TMG. For the other sector, generated by $\un\be$, using the mode decomposition representation $\be_1=e^{i m x^+}$ and $\be_2=e^{i n x^+}$, and calling $Q_{\un\be^1}[g]={{L}}_m$ and $Q_{\un\be^2}[g]={{L}}_n$:
\beq
{{L}}_m=\frac{1}{18 \pi  G \mu ^3 L^4}\int_0^{2\pi}\D \phi \ e^{imx^+}\Big(- 2GL\Delta(\mu ^4 L^4+18 \mu ^2 L^2
  h(x^+) +243h(x^+) ^2)+9 \mu ^2 L^2 f_{++}(x^+) \Big),
\eeq
we obtain
\beq
i\{{{L}}_m,{{L}}_n\}=(m-n) {{L}}_{m+n}+{c\over 12}m^3 \delta_{m+n,0}, \qquad c=\frac{\mu ^2 L^2+9}{3 G \mu}.
\eeq
Note that both central charges are sensitive to the sign of $\mu$. And finally the mixed sector, given our conventions, gives the usual semi-direct action of Virasoro on the $\hat u(1)$ algebra
\beq
i \{{{L}}_m,{{P}}_n\}=-n {{P}}_{m+n}.
\eeq

So, to summarize, the total symmetry group, also when $j_{++}$ varies, is
\beq
\text{Vir} \ltimes U(1).
\eeq
The total algebra is
\beqn
i\{{{L}}_m,{{L}}_n\}=(m-n) {{L}}_{m+n}+{c\over 12}m^3 \delta_{m+n,0},\quad
i \{{{L}}_m,{{P}}_n\} =-n {{P}}_{m+n},\quad
i\{{{P}}_m,{{P}}_n\} = m {k\over 2} \delta_{m+n,0},
\eeqn
with central extensions
\beqn
c = \frac{\mu ^2 L^2+9}{3 G \mu},\qquad
k = -sg(\Delta)\frac{8}{\mu L}.
\eeqn
Notice that, once the sign of $j_{++}$ given, both central extensions are field independent, in contrast with the previous case. Therefore, this bulk in these coordinates is dual to thermal states of a WCFT in the canonical ensemble. This is the reason why we denote the charges mode decomposition without the tilde, in line with (\ref{CSA}-\ref{CSAf}).

Finally, we can see how WBTZ black holes are described in this setup. Setting
\beqn
j_{++} = \frac{\mu ^2 L^2-9}{72 G L (L M-J)}, \quad
h= 0,\quad
f_{++} = \frac{1}{9} G L \left(\mu ^2 L^2+9\right) (J+L M),\quad
\Delta={\mu^2L^2-9 \over 144 G L j_{++}}= {LM-J\over 2},
\eeqn
we observe that, since the absence of naked singularities requires $LM\geq |J|$, the quantity $\Delta$ is always positive, so that $sg(\Delta)=1$ and $|\Delta|=\Delta$.
Therefore the charges in the canonical ensemble take the form
\beqn
{{P}}_m=\frac{2\sqrt{2(LM-J)}}{3}\delta_{m,0},\qquad
{{L}}_m=\Big(\frac{2 J \mu  L}{9}+\frac{J}{\mu  L}+\frac{M}{\mu }\Big)\delta_{m,0}.
\eeqn
The TMG mass and angular momentum of these solutions are defined as
\beqn
\pmb{M}  = Q_{\pa_t} = {1\over L}\Big(Q_{\pa_+}+Q_{\pa_-}\Big), \qquad
\pmb{J} = Q_{ \pa_\phi}= Q_{\pa_+}-Q_{\pa_-}.
\eeqn
As a consequence of the boundary conditions used here, the relationship between ${{P}}_0$ and $Q_{\pa_-}$ is non-trivial. In particular, since we have performed a field-dependent redefinition involving a multiplication rather than a linear shift, the charges are non-linearly related. Since $\pa_-$ is an exact Killing, we can compute separately this charge for WBTZ backgrounds, and then compare it with the results here.  We find\footnote{This relationship has an important extra factor of ${1\over 2}$ with respect to the naive linear rescaling. This comes about because of the field dependent redefinition. Note that it is only for WBTZ black holes that we can separately compute $Q_{\pa_-}$, which is otherwise non integrable.}
\beqn
Q_{\pa_-}={{P}}_0 {\mu L\sqrt{|\Delta|}\over 6},\qquad
Q_{\pa_+}={{L}}_0+{{P}}_0  {\mu L\sqrt{|\Delta|}\over 6},
\eeqn
which implies
\beqn
\pmb{M} = {1\over L}\Big({{L}}_0+{{P}}_0 {2\mu L\sqrt{|\Delta|}\over 3}\Big)={2\mu L M\over 9}+{LM+J\over \mu L^2},\qquad
\pmb{J} ={{L}}_0= {2\mu L J\over 9}+{LM+J\over \mu L}.
\eeqn
While the expression for the TMG mass and angular momentum in terms of the zero modes depends on the explicit realization of the algebra, their intrinsic value as charges is invariant. Indeed,
as non-trivial consistency check, we observe that $\pmb{M}=\tilde{M}$ and $\pmb{J}=\tilde J$, with $\tilde M$ and $\tilde J$ given in \eqref{tM}.

Recalling that for WBTZ black holes we have that $sg(\Delta)=1$, we obtain a relationship between the zero modes here and those in the quadratic ensemble of the previous section given by
\beqn
 \tilde{{L}}_0 = {{L}}_0+\frac{\mu L}{8} {{P}}_0^2={{L}}_0-{{{P}}_0^2\over k} , \qquad \tilde{{P}}_0 = {\mu L\over 8}{{P}}_0^2 = -{{{P}}_0^2\over k}, 
\eeqn
which exactly reproduces \eqref{Red0}.

To summarize, we have seen how two different restriction on the solution space leading to different boundary conditions can be used to make the charges integrable, yielding different realizations of the asymptotic symmetry algebra. With our new metric falloffs, the two methods give a bulk that is dual to a WCFT in either canonical or quadratic ensemble. We will focus in particular on the latter, whose modes are reported in \eqref{tpo}, because, as explained previously, the warped Cardy formula is well defined in this case. The novelty is that we do not reach this ensemble from a bulk change of coordinates, instead we found enhanced boundary conditions to automatically accommodate it. 

\section{Entropy matching}\label{s5}

We turn in this section to the matching between thermodynamic bulk entropy in TMG and boundary warped Cardy formula in quadratic ensemble. As is clear from \eqref{entropyQ}, this will require identifying the vacuum state on which the partition function projects in the Cardy regime.
Although we will only indirectly identify the vacuum via intuition gathered with limiting procedures, we discuss at the end of this section possible techniques to intrinsically identify it. These techniques require a deep understanding of the global structure of these solutions, which is currently under investigation and on which we will report later.

\subsection{Bulk thermodynamics}\label{sec4}

To study the thermodynamics of WBTZ solutions, following 
\cite{Kraus:2005zm, Solodukhin:2005ah, Tachikawa:2006sz, Bouchareb:2007yx, Detournay:2012ug}, we bring the line element to the ADM form:
\beq
\D s^2=-N(r)^2 \D t^2+{\D r^2\over f(r)^2}+R(r)^2(N^\phi(r)\D t+\D \phi)^2.\label{ADM}
\eeq
In the coordinates $(t,r,\phi)$, with $\varepsilon^{tr\phi}=1$, using $x^\pm={t\over L}\pm\phi$, the WBTZ metric  is given by\footnote{\label{comp}In comparing with \cite{Detournay:2015ysa}, one should set $G=1$, $L=1$.}
\beqn
g_{rr} &=&  \frac{L^2 r^2}{16 G^2 J^2 L^2-8 G L^2 M r^2+r^4} \label{WBTZgrr}, \\
 g_{tt}&=& -\frac{H^2 \left(4 G L (J-2 L M)+r^2\right)^2}{4 G L^3 (L M-J)}+8 G M-\frac{r^2}{L^2}, \\
 g_{t\phi}&=& 4 G J-\frac{H^2 \left(r^2-4 G J L\right) \left(4 G L (J-2 L M)+r^2\right)}{4 G L^2 (L M-J)}, \\
g_{\phi \phi} &=& r^2-\frac{H^2 \left(r^2-4 G J L\right)^2}{4 G L (L M-J)} \label{WBTZgphiphi},
\eeqn
which is exactly \eqref{WBTZ}.
Therefore, the ADM data read
\beqn
N(r)^2 &=& -\frac{4 G \left(2 H^2-1\right) (J-L M) \left(16 G^2 J^2 L^2-8 G L^2 M r^2+r^4\right)}{L \left(16 G^2 H^2
   J^2 L^2-4 G L r^2 \left(\left(2 H^2-1\right) J+L M\right)+H^2 r^4\right)},\\
f(r)^2 &=& \frac{16 G^2 J^2}{r^2}-8 G M+\frac{r^2}{L^2},\\
R(r)^2 &=& -\frac{16 G^2 H^2 J^2 L^2-4 G L r^2 \left(2 H^2 J-J+L M\right)+H^2 r^4}{4 G L (-J+L M)}, \\
N^{\phi}(r) &=& \frac{-16 G^2 J L^2 \left(\left(H^2-1\right) J+\left(1-2 H^2\right) L M\right)-8 G H^2 L^2 M r^2+H^2
   r^4}{L \left(16 G^2 H^2 J^2 L^2-4 G L r^2 \left(\left(2 H^2-1\right) J+L M\right)+H^2 r^4\right)}. 
\eeqn
Applying in particular the results of \cite{Bouchareb:2007yx}, the black hole entropy in TMG, given \eqref{ADM}, is
\beq
S^{TMG}_{\pm}={\pi\over 2 G}R(r_\pm)-{\pi\over 2\mu G}{R(r)^2f(r)N^\phi(r)^\prime\over 2 N(r)}\Big\vert_{r=r_\pm},\label{EntropyTMG}
\eeq
where $\mu$ is the Chern-Simons coupling and $r_\pm$ are the two positive roots, solutions of the equation $f(r)=0$ and are given by
\beqn
r_\pm = 2 \sqrt{G L} \sqrt{L M\pm\sqrt{L^2 M^2- J^2}},
\eeqn
for all values of $J$.
For WBTZ we thus gather the outer entropy
\beqn
S^{TMG}_{+} &=&\frac{\pi  L \left(\left(3-4 H^2\right) \sqrt{\sqrt{L^2 M^2-J^2}+L M}+\text{sg}(J) \sqrt{L M-\sqrt{L^2
   M^2-J^2}}\right)}{3 \sqrt{G \left(1-2 H^2\right) L}}\label{EntropyWBTZ},
\eeqn
where $sg$ is again the sign function. The presence of $sg(J)$ is due to the fact that we are dealing with a theory sensitive to parity, due to the presence of the Hodge-dual Cotton tensor in the equations of motion.

The WBTZ mass $\tilde{M}  = Q_{\pa_t} = \frac{1}{L} \left( {\tilde{{L}}}_0 + {\tilde{{P}}}_0 \right)$ and angular momentum $\tilde{J}= Q_{\pa_\phi}= {\tilde{{L}}}_0 - {\tilde{{P}}}_0 $  are given by \eqref{tM},
i.e.
\beqn
\tilde{M}&=&{2\mu L M\over 9}+{LM+J\over \mu L^2}= {\left(3-4 H^2\right) L M+J \over 3 \sqrt{1-2 H^2} L},\\
\tilde{J}&=&{2\mu L J\over 9}+{LM+J\over \mu L}= {\left(3-4 H^2\right) J+L M\over 3 \sqrt{1-2 H^2}}. 
\eeqn

As expected, the entropy \eqref{EntropyWBTZ} satisfies the first law
\beq
\D \tilde{M}=T \D S^{TMG}_+ + \Omega \D \tilde{J}, 
\eeq
with
\beq
T=\frac{r_{+}^2-r_{-}^2}{2 \pi r_{+} L^2}=\frac{2}{\pi} \sqrt{\frac{G \left(L^2 M^2-J^2\right)}{L^3 \left(\sqrt{L^2 M^2-J^2}+L M\right)}},
\eeq
and
\beq
\Omega=sg(J)\frac{r_{-}}{L r_{+}}=\frac{1}{L J} \sqrt{2 L M \left(L M-\sqrt{L^2 M^2-J^2}\right)-J^2}.
\eeq
Finally, we note that in the CSS limit we find the BTZ entropy in TMG, as found e.g. in \cite{Ciambelli:2020shy}. This is the standard derivation of the black hole thermodynamic entropy, we now turn our attention to the dual viewpoint.

\subsection{WCFT Cardy formula and matching}

In this section we would like to obtain the entropy from the counting of degeneracy of states in the boundary field theory.
The first important step is to define the vacuum of the theory. The usual approach is to use as a guideline the enhancement of isometries. We now compute the local Killing vectors preserving the WBTZ metric in the quadratic ensemble (\ref{WBTZgrr}-\ref{WBTZgphiphi}). These are $\pa_t$, $\pa_\phi$ and $\un \chi_{\pm}$ given by 

\beqn
\chi_{\pm}^r&=&e^{\pm 2 \sqrt{2}  \sqrt{\frac{G (J+L M)}{L}} \left(\frac{t}{L}+\phi \right)}  \frac{ \sqrt{16 G^2 J^2 L^2-8 G L^2 M r^2+r^4}}{2 r}, \\
\chi_{\pm}^t &=& \mp  e^{\pm 2 \sqrt{2}  \sqrt{\frac{G (J+L M)}{L}} \left(\frac{t}{L}+\phi \right)}  \frac{ L \left(4 G J L+r^2\right) }{4  \sqrt{2}   \sqrt{\frac{G (J+L M) \left(16 G^2 J^2 L^2-8 G L^2 M r^2+r^4\right)}{L}}} , \\
\chi_{\pm}^\phi &=&\pm  e^{\pm 2 \sqrt{2}  \sqrt{\frac{G (J+L M)}{L}} \left(\frac{t}{L}+\phi \right)} \frac{ \left(4 G L (J+2 L M)-r^2\right) }{ 4  \sqrt{2}   \sqrt{\frac{G (J+L M) \left(16 G^2 J^2 L^2-8 G L^2 M r^2+r^4\right)}{L}}} .
\eeqn 
These vectors, combined as $\{\frac{L}{2\sqrt{2} \sqrt{G L (J+L M)}}(L\pa_t+\pa_\phi),\un\chi_\pm\}$, satisfy an $SL(2,\mathbb{R})$ algebra. 

We observe that these local isometries are extended to hold globally if the exponential is $2\pi$ periodic in $\phi$. Thus, the WBTZ solution admits global isometries if and only if
\beq \label{vacuum condition}
2 \sqrt{2} \sqrt{\frac{G (J+L M)}{L}}= \pm i.
\eeq This equation therefore has solution $ M = -\frac{1}{8 G}-\frac{J}{L} $.  The metric given by (\ref{WBTZgrr}-\ref{WBTZgphiphi}) with $M$ taking this value therefore represents a family of solutions with enhanced global symmetries. A crucial observation is that the enhancement of symmetries does not single out the vacuum, contrary to what happens for BTZ black holes. Indeed, for the latter, we impose to enhance the symmetries to global $SL(2,\mathbb{R})_L \times SL(2,\mathbb{R})_R$, which imposes more restrictions on the metric parameters, allowing to immediately single out the unique vacuum.

We notice that inside the family of solutions with enhanced symmetry, one particular solution deserves special attention, the one with J = 0. In this case we obtain $M=-\frac{1}{8 G}$, and the line element is explicitly written
\beq
\D s^2_{\text{vac}}=   (L^2+r^2)(2H^2r^2+L^2(-1+2H^2)) {\D t^2\over L^4}+{L^2\D r^2\over (L^2+r^2)}+4H^2r^2(L^2+r^2){\D t\D \phi\over L^3}+(r^2+{2H^2r^4\over L^2})\D \phi^2.
\eeq

This metric differs from the AdS$_3$ global vacuum, but reduces to it for $H\to 0$ (i.e. $\mu \ell=3$ with $L=\ell$). We take this external input on the analysis to be the feature singling out the true vacuum of the theory. As already pointed out, we stress that this is not an intrinsic definition of the vacuum, but rather comes from a limiting procedure. We will return to this at the end of the section, where we present a more detailed discussion.
For this specific solution, the values of the TMG charges   are
\beqn
\tilde{M}=-{\left(3-4 H^2\right)  \over 24 \sqrt{1-2 H^2} G},\qquad
\tilde{J}= -{L\over 24 \sqrt{1-2 H^2}G}.
\eeqn
We remark that the TMG angular moment $\tilde J$ is non-vanishing on the vacuum solution, even though the metric parameter $J$, which corresponds to the Einstein angular momentum in the Einstein limit, is vanishing.

We want now to evaluate the WBTZ entropy \eqref{EntropyWBTZ}, which is reproduced by counting the degeneracy of states in the dual WCFT. In the quadratic ensemble, the Warped Cardy formula takes the form \eqref{entropyQ}:
\beq
S_{\mathrm{WCFT}}=4 \pi \sqrt{-{\tilde{{P}}}_{0}^{v a c}{\tilde{{P}}}_{0}}+4 \pi \sqrt{-{\tilde{{L}}}_{0}^{v a c}{\tilde{{L}}}_{0}} \label{SWCFT}.
\eeq 
 For the specific solution identified above, the zero modes (\ref{tpo}) become
\beqn
\tilde{{{P}}}_0^{\text{vac}}=-\frac{\sqrt{1-2 H^2} L}{24 G} \label{P0Vac},\qquad
\tilde{{{L}}}_0^{\text{vac}}=\frac{\left(H^2-1\right) L}{12 G \sqrt{1-2 H^2}} \label{L0Vac}.
\eeqn 
Plugging this and (\ref{tpo}) in \eqref{SWCFT}, and using $\mu L= 3 \sqrt{(1-2 H^2)}$, we find
\beq
S_{\mathrm{WCFT}}=\frac{1}{3} \sqrt{2} \pi  \left(\sqrt{\frac{\left(2 H^2-1\right) L (J-L
   M)}{G}}+2 \sqrt{-\frac{\left(H^2-1\right)^2 L (J+L M)}{G \left(2
   H^2-1\right)}}\right).
\eeq
After some non-trivial manipulations, this expression exactly matches the bulk thermodynamic WBTZ entropy \eqref{EntropyWBTZ}.

In conclusion, we return here to the discussion on the vacuum. To match the entropy, we used the specific solution whose charges coincide with the AdS$_3$ global vacuum in the limit $H\to 0$. We argue that this argument for singling out this vacuum must be improved with a more intrinsic and fundamental argument, lacking at present. There are two possible resolutions that we are currently exploring. First, it could be useful to introduce supersymmetry, and require that the true vacuum inside the family of solutions found is the one for which also all the supersymmetry generators are globally well-defined, which could potentially give further constraints. Indeed, it is known that certain BTZ and WAdS$_3$ black holes in canonical ensemble exhibit supersymmetry \cite{Coussaert:1993jp, Compere:2008cw}, and one could expect the vacuum state to display maximal supersymmetry. Second, a thorough analysis of the global structure of these solutions could give us a better understanding of the properties of this family of solutions, such as geodesic completeness, chronological and/or conical singularities, and the issue of closed time-like curves. 

\section{Conclusions and outlook}

We have introduced a consistent set of metric falloffs allowing to extract the asymptotic symmetry algebra directly in the quadratic ensemble. In this setup, asymptotic charges  are generically non integrable. Two possible boundary conditions arose, fixing part of the solution space or performing a field redefinition of the symmetry generators. These two procedures give an asymptotic symmetry algebra in quadratic and canonical ensemble, respectively. Using the former, we tested that we are in the regime of validity of the Cardy warped formula and show that the boundary counting of degeneracy of states correctly reproduces the bulk thermodynamic entropy for WBTZ black holes. To do so, we had to identify the vacuum of the theory. We showed that the enhancement of Killing isometries to globally well-defined symmetries is not enough, and propose the identification of the vacuum via a limiting procedure.

This project opens the door to both short-term and long-term investigations. The most pressing and natural continuation of this project is to find an intrinsic definition of the vacuum for the family of solutions with enhanced symmetry. As we discussed, we plan to introduce supersymmetry, to see if it imposes further constraints on the solutions. Another direction is to study in detail the global structure of these solutions. This might shed light on their topological properties. We would like to find a complete Penrose-Carter diagram of these solutions, and study their inextensibility. Another direction to pursue is the question of integrability of charges. We confirmed here that it seems always possible to render charges integrable by redefining the symmetry generators by introducing field/ state dependence, as has also been carried out in  \cite{Afshar:2016wfy, Perez:2016vqo, Grumiller:2019fmp,  Adami:2021nnf}. Another mechanism to make charges integrable, recently proposed in \cite{Ciambelli:2021nmv} (see also \cite{Freidel:2021dxw}), is to carefully treat embeddings. It would be interesting to apply this new mechanism to our specific construction here. Other questions stemming from the covariant phase-space formalism, when applied to TMG, are also worth investigating, such as the classification of ambiguities.
On the long term, there are some fascinating directions to pursue. First, we here restricted our analysis to $0< \mu L< 6$. At $\mu L=6$, the cosmological constant vanishes, whereas for $\mu L>6$ the metric has positive cosmological constant. Our analysis of  surface charges is not expected to break down, allowing us to sail from negative to positive cosmological constants, an idea already explored in \cite{Anninos:2009jt}. Secondly, we saw that in our analysis we restricted the would-be boundary metric to have non-varying conformal factor. This restriction could be lifted, in the spirit of \cite{Alessio:2020ioh}. A similar construction  appeared recently in \cite{Chaturvedi:2020jyy}. Eventually, we expect to have some more general boundary conditions arising at specific values of the Chern-Simons coupling. We briefly touch upon this for $\mu L=3$ in Appendix \ref{AppA}. In this setup, we expect to be able to make contact with \cite{Henneaux:2010fy}. A far-reaching consequence could be to shed light on the still elusive properties of logarithmic CFTs \cite{Henneaux:2009pw,Grumiller:2008qz, Giribet:2008bw, Grumiller:2013at}, and study the possibility of constructing their warped version.

\vspace{0.5cm}

\paragraph{Acknowledgments}

We thank Marc Henneaux for valuable discussions.
 AA is a Research Fellow of the Fonds de la Recherche Scientifique F.R.S.-FNRS (Belgium). The research of LC was partially supported by a Marina Solvay 
Fellowship, by the ERC Advanced Grant ``High-Spin-Grav" and by 
FNRS-Belgium (convention FRFC PDR T.1025.14 and convention IISN 4.4503.15).  SD is a Research Associate of the Fonds de la Recherche Scientifique F.R.S.-FNRS (Belgium). SD was supported in part by IISN -- Belgium (convention 4.4503.15) and benefited from the support of the Solvay Family. SD acknowledges support of the Fonds de la Recherche Scientifique F.R.S.-FNRS (Belgium) through the CDR project C 60/5 - CDR/OL ``Horizon holography : black holes and field theories" (2020-2022).  This research was supported in part by the National Science Foundation under Grant No. NSF PHY-1748958.  
The work of AS is supported by  the ``Fonds pour la Formation et la Recherche dans l'Industrie et dans l'Agriculture", FRIA (Belgium).

\appendix
\section{New boundary conditions including pp waves}\label{AppA}

At $\mu={3\over L}={3\over \ell}$, (where $\ell={1\over\sqrt{|\Lambda|}}$), the solutions (\ref{gpp}-\ref{gmm}) can be generalized to allow $j_{++}$ to be an arbitrary function $j_{++}(x^+)$. The resulting class of metric reads
\beqn
\D s^2&=&(\D x^+)^2 \left(f_{++}(x^+)+r^2 h(x^+)+j_{++}(x^+) r^4\right)+ \D x^- \D x^+ (2 j_{+-}(x^+)+1)r^2+\D r^2\frac{\ell^2 }{r^2}.
\eeqn
In fact, these solutions can be generalized for arbitrary values of positive TMG coupling, $\mu>0$, as follows
\beqn
\D s^2&=&(\D x^+)^2 \left(f_{++}(x^+)+r^2 h(x^+)+j_{++}(x^+) r^{\mu\ell +1}\right)+ \D x^-  \D x^+ (2 j_{+-}(x^+)+1)r^2+\D r^2\frac{\ell^2 }{r^2},
\eeqn
where $j_{+-}(x^+)\leq -{1\over 2}$.  
This class of solutions includes the pp-waves. The asymptotic symmetry group is again Vir$\ltimes$ U(1). 

For $\mu\ell<1$, these boundary conditions generalize CSS boundary conditions \cite{Compere:2013bya} in TMG in the same way \cite{Henneaux:2010fy}  generalizes Brown-Hennaux boundary conditions \cite {Brown:1986nw} in TMG. For $\mu\ell\geq 3$, these generalize the null warped and null z-warped boundary conditions of \cite{Anninos:2010pm}.  The details about the charges and other features of these solutions will be presented elsewhere.
We also plan to explore their relationship with the boundary conditions involving non-integer ($\mu$-dependent) powers of $r$ and $\log r$ introduced in \cite{Henneaux:2010fy} (see also \cite{Henneaux:2009pw, Grumiller:2008qz, Giribet:2008bw, Grumiller:2013at}).

\providecommand{\href}[2]{#2}\begingroup\raggedright\endgroup


\begin{thebibliography}{10}

\bibitem{Teitelboim:1983ux}
C.~Teitelboim, ``{Gravitation and Hamiltonian Structure in Two Space-Time
  Dimensions},'' \href{http://dx.doi.org/10.1016/0370-2693(83)90012-6}{{\em
  Phys. Lett. B} {\bf 126} (1983)  41--45}.

\bibitem{Deser:1983tn}
S.~Deser, R.~Jackiw, and G.~'t~Hooft, ``{Three-Dimensional Einstein Gravity:
  Dynamics of Flat Space},''
  \href{http://dx.doi.org/10.1016/0003-4916(84)90085-X}{{\em Annals Phys.} {\bf
  152} (1984)  220}.

\bibitem{Deser:1983nh}
S.~Deser and R.~Jackiw, ``{Three-Dimensional Cosmological Gravity: Dynamics of
  Constant Curvature},''
  \href{http://dx.doi.org/10.1016/0003-4916(84)90025-3}{{\em Annals Phys.} {\bf
  153} (1984)  405--416}.

\bibitem{Jackiw:1984je}
R.~Jackiw, ``{Lower Dimensional Gravity},''
  \href{http://dx.doi.org/10.1016/0550-3213(85)90448-1}{{\em Nucl. Phys. B}
  {\bf 252} (1985)  343--356}.

\bibitem{doi:10.1142/0622}
J.~D. Brown, ``Lower dimensional gravity,''
  \href{http://arxiv.org/abs/https://www.worldscientific.com/doi/pdf/10.1142/0622}{{\tt
  https://www.worldscientific.com/doi/pdf/10.1142/0622}}.

\bibitem{doi:10.1142/2295}
R.~Jackiw, ``Diverse topics in theoretical and mathematical physics,''
  \href{http://arxiv.org/abs/https://www.worldscientific.com/doi/pdf/10.1142/2295}{{\tt
  https://www.worldscientific.com/doi/pdf/10.1142/2295}}.

\bibitem{Carlip:1995zj}
S.~Carlip, ``{Lectures on (2+1) dimensional gravity},'' {\em J. Korean Phys.
  Soc.} {\bf 28} (1995)  S447--S467,
  \href{http://arxiv.org/abs/gr-qc/9503024}{{\tt arXiv:gr-qc/9503024}}.

\bibitem{Brown:1986nw}
J.~D. Brown and M.~Henneaux, ``{Central Charges in the Canonical Realization of
  Asymptotic Symmetries: An Example from Three-Dimensional Gravity},''
  \href{http://dx.doi.org/10.1007/BF01211590}{{\em Commun. Math. Phys.} {\bf
  104} (1986)  207--226}.

\bibitem{Banados:1992wn}
M.~Banados, C.~Teitelboim, and J.~Zanelli, ``{The Black hole in
  three-dimensional space-time},''
  \href{http://dx.doi.org/10.1103/PhysRevLett.69.1849}{{\em Phys. Rev. Lett.}
  {\bf 69} (1992)  1849--1851}, \href{http://arxiv.org/abs/hep-th/9204099}{{\tt
  arXiv:hep-th/9204099}}.

\bibitem{Banados:1992gq}
M.~Banados, M.~Henneaux, C.~Teitelboim, and J.~Zanelli, ``{Geometry of the
  (2+1) black hole},'' \href{http://dx.doi.org/10.1103/PhysRevD.48.1506}{{\em
  Phys. Rev. D} {\bf 48} (1993)  1506--1525},
  \href{http://arxiv.org/abs/gr-qc/9302012}{{\tt arXiv:gr-qc/9302012}}.
  [Erratum: Phys.Rev.D 88, 069902 (2013)].

\bibitem{Guica:2008mu}
M.~Guica, T.~Hartman, W.~Song, and A.~Strominger, ``{The Kerr/CFT
  Correspondence},'' \href{http://dx.doi.org/10.1103/PhysRevD.80.124008}{{\em
  Phys. Rev. D} {\bf 80} (2009)  124008},
  \href{http://arxiv.org/abs/0809.4266}{{\tt arXiv:0809.4266 [hep-th]}}.

\bibitem{El-Showk:2011euy}
S.~El-Showk and M.~Guica, ``{Kerr/CFT, dipole theories and nonrelativistic
  CFTs},'' \href{http://dx.doi.org/10.1007/JHEP12(2012)009}{{\em JHEP} {\bf 12}
  (2012)  009}, \href{http://arxiv.org/abs/1108.6091}{{\tt arXiv:1108.6091
  [hep-th]}}.

\bibitem{Compere:2012jk}
G.~Comp\`ere, ``{The Kerr/CFT correspondence and its extensions},''
  \href{http://dx.doi.org/10.1007/s41114-017-0003-2}{{\em Living Rev. Rel.}
  {\bf 15} (2012)  11}, \href{http://arxiv.org/abs/1203.3561}{{\tt
  arXiv:1203.3561 [hep-th]}}.

\bibitem{Detournay:2012pc}
S.~Detournay, T.~Hartman, and D.~M. Hofman, ``{Warped Conformal Field
  Theory},'' \href{http://dx.doi.org/10.1103/PhysRevD.86.124018}{{\em Phys.
  Rev. D} {\bf 86} (2012)  124018}, \href{http://arxiv.org/abs/1210.0539}{{\tt
  arXiv:1210.0539 [hep-th]}}.

\bibitem{Guica:2017lia}
M.~Guica, ``{An integrable Lorentz-breaking deformation of two-dimensional
  CFTs},'' \href{http://dx.doi.org/10.21468/SciPostPhys.5.5.048}{{\em SciPost
  Phys.} {\bf 5} (2018) no.~5, 048},
  \href{http://arxiv.org/abs/1710.08415}{{\tt arXiv:1710.08415 [hep-th]}}.

\bibitem{Haco:2018ske}
S.~Haco, S.~W. Hawking, M.~J. Perry, and A.~Strominger, ``{Black Hole Entropy
  and Soft Hair},'' \href{http://dx.doi.org/10.1007/JHEP12(2018)098}{{\em JHEP}
  {\bf 12} (2018)  098}, \href{http://arxiv.org/abs/1810.01847}{{\tt
  arXiv:1810.01847 [hep-th]}}.

\bibitem{Aggarwal:2019iay}
A.~Aggarwal, A.~Castro, and S.~Detournay, ``{Warped Symmetries of the Kerr
  Black Hole},'' \href{http://dx.doi.org/10.1007/JHEP01(2020)016}{{\em JHEP}
  {\bf 01} (2020)  016}, \href{http://arxiv.org/abs/1909.03137}{{\tt
  arXiv:1909.03137 [hep-th]}}.

\bibitem{Rooman:1998xf}
M.~Rooman and P.~Spindel, ``{Godel metric as a squashed anti-de Sitter
  geometry},'' \href{http://dx.doi.org/10.1088/0264-9381/15/10/024}{{\em Class.
  Quant. Grav.} {\bf 15} (1998)  3241--3249},
  \href{http://arxiv.org/abs/gr-qc/9804027}{{\tt arXiv:gr-qc/9804027}}.

\bibitem{Duff:1998cr}
M.~J. Duff, H.~Lu, and C.~N. Pope, ``{AdS(3) x S**3 (un)twisted and squashed,
  and an O(2,2,Z) multiplet of dyonic strings},''
  \href{http://dx.doi.org/10.1016/S0550-3213(98)00810-4}{{\em Nucl. Phys. B}
  {\bf 544} (1999)  145--180}, \href{http://arxiv.org/abs/hep-th/9807173}{{\tt
  arXiv:hep-th/9807173}}.

\bibitem{Israel:2003ry}
D.~Israel, C.~Kounnas, and M.~P. Petropoulos, ``{Superstrings on NS5
  backgrounds, deformed AdS(3) and holography},''
  \href{http://dx.doi.org/10.1088/1126-6708/2003/10/028}{{\em JHEP} {\bf 10}
  (2003)  028}, \href{http://arxiv.org/abs/hep-th/0306053}{{\tt
  arXiv:hep-th/0306053}}.

\bibitem{Israel:2004vv}
D.~Israel, C.~Kounnas, D.~Orlando, and P.~M. Petropoulos, ``{Electric/magnetic
  deformations of S**3 and AdS(3), and geometric cosets},''
  \href{http://dx.doi.org/10.1002/prop.200410190}{{\em Fortsch. Phys.} {\bf 53}
  (2005)  73--104}, \href{http://arxiv.org/abs/hep-th/0405213}{{\tt
  arXiv:hep-th/0405213}}.

\bibitem{Bouchareb:2007yx}
A.~Bouchareb and G.~Clement, ``{Black hole mass and angular momentum in
  topologically massive gravity},''
  \href{http://dx.doi.org/10.1088/0264-9381/24/22/018}{{\em Class. Quant.
  Grav.} {\bf 24} (2007)  5581--5594},
  \href{http://arxiv.org/abs/0706.0263}{{\tt arXiv:0706.0263 [gr-qc]}}.

\bibitem{Carlip:2008eq}
S.~Carlip, S.~Deser, A.~Waldron, and D.~K. Wise, ``{Topologically Massive AdS
  Gravity},'' \href{http://dx.doi.org/10.1016/j.physletb.2008.07.057}{{\em
  Phys. Lett. B} {\bf 666} (2008)  272--276},
  \href{http://arxiv.org/abs/0807.0486}{{\tt arXiv:0807.0486 [hep-th]}}.

\bibitem{Anninos:2008fx}
D.~Anninos, W.~Li, M.~Padi, W.~Song, and A.~Strominger, ``{Warped AdS(3) Black
  Holes},'' \href{http://dx.doi.org/10.1088/1126-6708/2009/03/130}{{\em JHEP}
  {\bf 03} (2009)  130}, \href{http://arxiv.org/abs/0807.3040}{{\tt
  arXiv:0807.3040 [hep-th]}}.

\bibitem{Compere:2008cv}
G.~Compere and S.~Detournay, ``{Semi-classical central charge in topologically
  massive gravity},''
  \href{http://dx.doi.org/10.1088/0264-9381/26/1/012001}{{\em Class. Quant.
  Grav.} {\bf 26} (2009)  012001}, \href{http://arxiv.org/abs/0808.1911}{{\tt
  arXiv:0808.1911 [hep-th]}}. [Erratum: Class.Quant.Grav. 26, 139801 (2009)].

\bibitem{Henneaux:2010fy}
M.~Henneaux, C.~Martinez, and R.~Troncoso, ``{More on Asymptotically Anti-de
  Sitter Spaces in Topologically Massive Gravity},''
  \href{http://dx.doi.org/10.1103/PhysRevD.82.064038}{{\em Phys. Rev. D} {\bf
  82} (2010)  064038}, \href{http://arxiv.org/abs/1006.0273}{{\tt
  arXiv:1006.0273 [hep-th]}}.

\bibitem{Bardeen:1999px}
J.~M. Bardeen and G.~T. Horowitz, ``{The Extreme Kerr throat geometry: A Vacuum
  analog of AdS(2) x S**2},''
  \href{http://dx.doi.org/10.1103/PhysRevD.60.104030}{{\em Phys. Rev. D} {\bf
  60} (1999)  104030}, \href{http://arxiv.org/abs/hep-th/9905099}{{\tt
  arXiv:hep-th/9905099}}.

\bibitem{Jugeau:2010nq}
F.~Jugeau, G.~Moutsopoulos, and P.~Ritter, ``{From accelerating and Poincare
  coordinates to black holes in spacelike warped AdS$_3$, and back},''
  \href{http://dx.doi.org/10.1088/0264-9381/28/3/035001}{{\em Class. Quant.
  Grav.} {\bf 28} (2011)  035001}, \href{http://arxiv.org/abs/1007.1961}{{\tt
  arXiv:1007.1961 [hep-th]}}.

\bibitem{Compere:2009zj}
G.~Compere and S.~Detournay, ``{Boundary conditions for spacelike and timelike
  warped $AdS_{3}$ spaces in topologically massive gravity},''
  \href{http://dx.doi.org/10.1088/1126-6708/2009/08/092}{{\em JHEP} {\bf 08}
  (2009)  092}, \href{http://arxiv.org/abs/0906.1243}{{\tt arXiv:0906.1243
  [hep-th]}}.

\bibitem{Blagojevic:2009ek}
M.~Blagojevic and B.~Cvetkovic, ``{Asymptotic structure of topologically
  massive gravity in spacelike stretched AdS sector},''
  \href{http://dx.doi.org/10.1088/1126-6708/2009/09/006}{{\em JHEP} {\bf 09}
  (2009)  006}, \href{http://arxiv.org/abs/0907.0950}{{\tt arXiv:0907.0950
  [gr-qc]}}.

\bibitem{Henneaux:2011hv}
M.~Henneaux, C.~Martinez, and R.~Troncoso, ``{Asymptotically warped anti-de
  Sitter spacetimes in topologically massive gravity},''
  \href{http://dx.doi.org/10.1103/PhysRevD.84.124016}{{\em Phys. Rev. D} {\bf
  84} (2011)  124016}, \href{http://arxiv.org/abs/1108.2841}{{\tt
  arXiv:1108.2841 [hep-th]}}.

\bibitem{Hofman:2011zj}
D.~M. Hofman and A.~Strominger, ``{Chiral Scale and Conformal Invariance in 2D
  Quantum Field Theory},''
  \href{http://dx.doi.org/10.1103/PhysRevLett.107.161601}{{\em Phys. Rev.
  Lett.} {\bf 107} (2011)  161601}, \href{http://arxiv.org/abs/1107.2917}{{\tt
  arXiv:1107.2917 [hep-th]}}.

\bibitem{Apolo:2018eky}
L.~Apolo and W.~Song, ``{Bootstrapping holographic warped CFTs or: how I
  learned to stop worrying and tolerate negative norms},''
  \href{http://dx.doi.org/10.1007/JHEP07(2018)112}{{\em JHEP} {\bf 07} (2018)
  112}, \href{http://arxiv.org/abs/1804.10525}{{\tt arXiv:1804.10525
  [hep-th]}}.

\bibitem{Castro:2015uaa}
A.~Castro, D.~M. Hofman, and G.~S\'arosi, ``{Warped Weyl fermion partition
  functions},'' \href{http://dx.doi.org/10.1007/JHEP11(2015)129}{{\em JHEP}
  {\bf 11} (2015)  129}, \href{http://arxiv.org/abs/1508.06302}{{\tt
  arXiv:1508.06302 [hep-th]}}.

\bibitem{Castro:2015csg}
A.~Castro, D.~M. Hofman, and N.~Iqbal, ``{Entanglement Entropy in Warped
  Conformal Field Theories},''
  \href{http://dx.doi.org/10.1007/JHEP02(2016)033}{{\em JHEP} {\bf 02} (2016)
  033}, \href{http://arxiv.org/abs/1511.00707}{{\tt arXiv:1511.00707
  [hep-th]}}.

\bibitem{Song:2016gtd}
W.~Song, Q.~Wen, and J.~Xu, ``{Modifications to Holographic Entanglement
  Entropy in Warped CFT},''
  \href{http://dx.doi.org/10.1007/JHEP02(2017)067}{{\em JHEP} {\bf 02} (2017)
  067}, \href{http://arxiv.org/abs/1610.00727}{{\tt arXiv:1610.00727
  [hep-th]}}.

\bibitem{Song:2016pwx}
W.~Song, Q.~Wen, and J.~Xu, ``{Generalized Gravitational Entropy for Warped
  Anti\textendash{}de Sitter Space},''
  \href{http://dx.doi.org/10.1103/PhysRevLett.117.011602}{{\em Phys. Rev.
  Lett.} {\bf 117} (2016) no.~1, 011602},
  \href{http://arxiv.org/abs/1601.02634}{{\tt arXiv:1601.02634 [hep-th]}}.

\bibitem{Compere:2013aya}
G.~Comp\`ere, W.~Song, and A.~Strominger, ``{Chiral Liouville Gravity},''
  \href{http://dx.doi.org/10.1007/JHEP05(2013)154}{{\em JHEP} {\bf 05} (2013)
  154}, \href{http://arxiv.org/abs/1303.2660}{{\tt arXiv:1303.2660 [hep-th]}}.

\bibitem{Hofman:2014loa}
D.~M. Hofman and B.~Rollier, ``{Warped Conformal Field Theory as Lower Spin
  Gravity},'' \href{http://dx.doi.org/10.1016/j.nuclphysb.2015.05.011}{{\em
  Nucl. Phys. B} {\bf 897} (2015)  1--38},
  \href{http://arxiv.org/abs/1411.0672}{{\tt arXiv:1411.0672 [hep-th]}}.

\bibitem{Jensen:2017tnb}
K.~Jensen, ``{Locality and anomalies in warped conformal field theory},''
  \href{http://dx.doi.org/10.1007/JHEP12(2017)111}{{\em JHEP} {\bf 12} (2017)
  111}, \href{http://arxiv.org/abs/1710.11626}{{\tt arXiv:1710.11626
  [hep-th]}}.

\bibitem{Compere:2013bya}
G.~Comp\`ere, W.~Song, and A.~Strominger, ``{New Boundary Conditions for
  AdS3},'' \href{http://dx.doi.org/10.1007/JHEP05(2013)152}{{\em JHEP} {\bf 05}
  (2013)  152}, \href{http://arxiv.org/abs/1303.2662}{{\tt arXiv:1303.2662
  [hep-th]}}.

\bibitem{DESER1982372}
S.~Deser, R.~Jackiw, and S.~Templeton, ``Topologically massive gauge
  theories,''
  \href{http://dx.doi.org/https://doi.org/10.1016/0003-4916(82)90164-6}{{\em
  Annals of Physics} {\bf 140} (1982) no.~2, 372--411}.
  \url{https://www.sciencedirect.com/science/article/pii/0003491682901646}.
  
  \bibitem{Detournay:2015ysa}
S.~Detournay and C.~Zwikel, ``{Phase transitions in warped AdS$_{3}$
  gravity},'' \href{http://dx.doi.org/10.1007/JHEP05(2015)074}{{\em JHEP} {\bf
  05} (2015)  074}, \href{http://arxiv.org/abs/1504.00827}{{\tt
  arXiv:1504.00827 [hep-th]}}.

\bibitem{Maldacena:1998bw}
J.~M. Maldacena and A.~Strominger, ``{AdS(3) black holes and a stringy
  exclusion principle},''
  \href{http://dx.doi.org/10.1088/1126-6708/1998/12/005}{{\em JHEP} {\bf 12}
  (1998)  005}, \href{http://arxiv.org/abs/hep-th/9804085}{{\tt
  arXiv:hep-th/9804085}}.

\bibitem{Ciambelli:2020shy}
L.~Ciambelli, S.~Detournay, and A.~Somerhausen, ``{New chiral gravity},''
  \href{http://dx.doi.org/10.1103/PhysRevD.102.106017}{{\em Phys. Rev. D} {\bf
  102} (2020) no.~10, 106017}, \href{http://arxiv.org/abs/2008.06793}{{\tt
  arXiv:2008.06793 [hep-th]}}.

\bibitem{Alessio:2020ioh}
F.~Alessio, G.~Barnich, L.~Ciambelli, P.~Mao, and R.~Ruzziconi, ``{Weyl charges
  in asymptotically locally AdS$_3$ spacetimes},''
  \href{http://dx.doi.org/10.1103/PhysRevD.103.046003}{{\em Phys. Rev. D} {\bf
  103} (2021) no.~4, 046003}, \href{http://arxiv.org/abs/2010.15452}{{\tt
  arXiv:2010.15452 [hep-th]}}.
  
  \bibitem{Henneaux:2009pw}
M.~Henneaux, C.~Martinez and R.~Troncoso,
``Asymptotically anti-de Sitter spacetimes in topologically massive gravity,''
\href{http://dx.doi.org/10.1103/PhysRevD.79.081502}{Phys. Rev. D \textbf{79}, 081502 (2009)}, \href{http://arxiv.org/abs/0901.2874}{{\tt arXiv:0901.2874 [hep-th]}}.

\bibitem{Anninos:2009jt}
D.~Anninos, ``{Sailing from Warped AdS(3) to Warped dS(3) in Topologically
  Massive Gravity},'' \href{http://dx.doi.org/10.1007/JHEP02(2010)046}{{\em
  JHEP} {\bf 02} (2010)  046}, \href{http://arxiv.org/abs/0906.1819}{{\tt
  arXiv:0906.1819 [hep-th]}}.

\bibitem{Anninos:2010pm}
D.~Anninos, G.~Compere, S.~de~Buyl, S.~Detournay, and M.~Guica, ``{The Curious
  Case of Null Warped Space},''
  \href{http://dx.doi.org/10.1007/JHEP11(2010)119}{{\em JHEP} {\bf 11} (2010)
  119}, \href{http://arxiv.org/abs/1005.4072}{{\tt arXiv:1005.4072 [hep-th]}}.

\bibitem{Ciambelli:2020eba}
L.~Ciambelli, C.~Marteau, P.~M. Petropoulos, and R.~Ruzziconi, ``{Gauges in
  Three-Dimensional Gravity and Holographic Fluids},''
  \href{http://dx.doi.org/10.1007/JHEP11(2020)092}{{\em JHEP} {\bf 11} (2020)
  092}, \href{http://arxiv.org/abs/2006.10082}{{\tt arXiv:2006.10082
  [hep-th]}}.

\bibitem{Ciambelli:2020ftk}
L.~Ciambelli, C.~Marteau, P.~M. Petropoulos, and R.~Ruzziconi,
  ``{Fefferman-Graham and Bondi Gauges in the Fluid/Gravity Correspondence},''
  \href{http://dx.doi.org/10.22323/1.376.0154}{{\em PoS} {\bf CORFU2019} (2020)
   154}, \href{http://arxiv.org/abs/2006.10083}{{\tt arXiv:2006.10083
  [hep-th]}}.

\bibitem{Chaturvedi:2020jyy}
P.~Chaturvedi, I.~Papadimitriou, W.~Song, and B.~Yu, ``{AdS$_3$ gravity and the
  complex SYK models},'' \href{http://arxiv.org/abs/2011.10001}{{\tt
  arXiv:2011.10001 [hep-th]}}.

\bibitem{Barnich:2010eb}
G.~Barnich and C.~Troessaert, ``{Aspects of the BMS/CFT correspondence},''
  \href{http://dx.doi.org/10.1007/JHEP05(2010)062}{{\em JHEP} {\bf 05} (2010)
  062}, \href{http://arxiv.org/abs/1001.1541}{{\tt arXiv:1001.1541 [hep-th]}}.

\bibitem{Wald:1993nt}
R.~M. Wald, ``{Black hole entropy is the Noether charge},''
  \href{http://dx.doi.org/10.1103/PhysRevD.48.R3427}{{\em Phys. Rev. D} {\bf
  48} (1993) no.~8, R3427--R3431},
  \href{http://arxiv.org/abs/gr-qc/9307038}{{\tt arXiv:gr-qc/9307038}}.

\bibitem{Iyer:1994ys}
V.~Iyer and R.~M. Wald, ``{Some properties of Noether charge and a proposal for
  dynamical black hole entropy},''
  \href{http://dx.doi.org/10.1103/PhysRevD.50.846}{{\em Phys. Rev. D} {\bf 50}
  (1994)  846--864}, \href{http://arxiv.org/abs/gr-qc/9403028}{{\tt
  arXiv:gr-qc/9403028}}.
  
  \bibitem{code}
G. Comp\`ere, package \textit{SurfaceCharges}, \ \href{http://www.ulb.ac.be/sciences/ptm/pmif/gcompere/package.html}{SurfaceCharges}

\bibitem{Adami:2020ugu}
H.~Adami, M.~M. Sheikh-Jabbari, V.~Taghiloo, H.~Yavartanoo, and C.~Zwikel,
  ``{Symmetries at null boundaries: two and three dimensional gravity cases},''
  \href{http://dx.doi.org/10.1007/JHEP10(2020)107}{{\em JHEP} {\bf 10} (2020)
  107}, \href{http://arxiv.org/abs/2007.12759}{{\tt arXiv:2007.12759
  [hep-th]}}.

\bibitem{Ruzziconi:2020wrb}
R.~Ruzziconi and C.~Zwikel, ``{Conservation and Integrability in
  Lower-Dimensional Gravity},'' \href{http://arxiv.org/abs/2012.03961}{{\tt
  arXiv:2012.03961 [hep-th]}}.
  
    
\bibitem{Adami:2021sko}
H.~Adami, M.~M.~Sheikh-Jabbari, V.~Taghiloo, H.~Yavartanoo and C.~Zwikel,
``{Chiral Massive News: Null Boundary Symmetries in Topologically Massive Gravity},''
 \href{http://dx.doi.org/10.1007/JHEP05(2021)261}{{\em JHEP} {\bf 05} (2021)
  261}, \href{https://arxiv.org/abs/2104.03992}{{\tt arXiv:2104.03992
  [hep-th]}}.
  
  \bibitem{Adami:2021nnf}
H.~Adami, D.~Grumiller, M.~M. Sheikh-Jabbari, V.~Taghiloo, H.~Yavartanoo, and
C.~Zwikel, ``{Null boundary phase space: slicings, news and memory},''
\href{http://arxiv.org/abs/2110.04218}{{\tt arXiv:2110.04218 [hep-th]}}.


\bibitem{Kraus:2005zm}
P.~Kraus and F.~Larsen, ``{Holographic gravitational anomalies},''
  \href{http://dx.doi.org/10.1088/1126-6708/2006/01/022}{{\em JHEP} {\bf 01}
  (2006)  022}, \href{http://arxiv.org/abs/hep-th/0508218}{{\tt
  arXiv:hep-th/0508218}}.

\bibitem{Solodukhin:2005ah}
S.~N. Solodukhin, ``{Holography with gravitational Chern-Simons},''
  \href{http://dx.doi.org/10.1103/PhysRevD.74.024015}{{\em Phys. Rev. D} {\bf
  74} (2006)  024015}, \href{http://arxiv.org/abs/hep-th/0509148}{{\tt
  arXiv:hep-th/0509148}}.

\bibitem{Tachikawa:2006sz}
Y.~Tachikawa, ``{Black hole entropy in the presence of Chern-Simons terms},''
  \href{http://dx.doi.org/10.1088/0264-9381/24/3/014}{{\em Class. Quant. Grav.}
  {\bf 24} (2007)  737--744}, \href{http://arxiv.org/abs/hep-th/0611141}{{\tt
  arXiv:hep-th/0611141}}.

\bibitem{Detournay:2012ug}
S.~Detournay, ``{Inner Mechanics of 3d Black Holes},''
  \href{http://dx.doi.org/10.1103/PhysRevLett.109.031101}{{\em Phys. Rev.
  Lett.} {\bf 109} (2012)  031101}, \href{http://arxiv.org/abs/1204.6088}{{\tt
  arXiv:1204.6088 [hep-th]}}.

\bibitem{Coussaert:1993jp}
O.~Coussaert and M.~Henneaux, ``{Supersymmetry of the (2+1) black holes},''
  \href{http://dx.doi.org/10.1103/PhysRevLett.72.183}{{\em Phys. Rev. Lett.}
  {\bf 72} (1994)  183--186}, \href{http://arxiv.org/abs/hep-th/9310194}{{\tt
  arXiv:hep-th/9310194}}.

\bibitem{Compere:2008cw}
G.~Compere, S.~Detournay, and M.~Romo, ``{Supersymmetric Godel and warped black
  holes in string theory},''
  \href{http://dx.doi.org/10.1103/PhysRevD.78.104030}{{\em Phys. Rev. D} {\bf
  78} (2008)  104030}, \href{http://arxiv.org/abs/0808.1912}{{\tt
  arXiv:0808.1912 [hep-th]}}.
  
  \bibitem{Afshar:2016wfy}
H.~Afshar, S.~Detournay, D.~Grumiller, W.~Merbis, A.~Perez, D.~Tempo and R.~Troncoso,
``Soft Heisenberg hair on black holes in three dimensions,''
\href{http://dx.doi.org/10.1103/PhysRevD.93.101503}{\tt{Phys. Rev. D \textbf{93}, no.10, 101503 (2016)}}
\href{https://arxiv.org/abs/1603.04824}{\tt{[arXiv:1603.04824 [hep-th]]}}.


\bibitem{Perez:2016vqo}
A.~P\'erez, D.~Tempo and R.~Troncoso,
``Boundary conditions for General Relativity on AdS$_{3}$ and the KdV hierarchy,''
\href{http://dx.doi.org/10.1007/JHEP06(2016)103}{JHEP \textbf{06}, 103 (2016)}, \href{http://arxiv.org/abs/1605.04490}{{\tt
		arXiv:1605.04490 [hep-th]}}.

\bibitem{Grumiller:2019fmp}
D.~Grumiller, A.~P\'erez, M.~M.~Sheikh-Jabbari, R.~Troncoso and C.~Zwikel,
``Spacetime structure near generic horizons and soft hair,''
\href{http://dx.doi.org/10.1103/PhysRevLett.124.041601}{{Phys. Rev. Lett. \textbf{124}, no.4, 041601 (2020)}}, \href{http://arxiv.org/abs/1908.09833}{{\tt
	[arXiv:1908.09833 [hep-th]]}}.
	
\bibitem{Ciambelli:2021nmv}
L.~Ciambelli, R.~G. Leigh, and P.-C. Pai, ``{Embeddings and Integrable Charges
  for Extended Corner Symmetry},'' \href{http://arxiv.org/abs/2111.13181}{{\tt
  arXiv:2111.13181 [hep-th]}}.

\bibitem{Freidel:2021dxw}
L.~Freidel, ``{A canonical bracket for open gravitational system},''
  \href{http://arxiv.org/abs/2111.14747}{{\tt arXiv:2111.14747 [hep-th]}}.

\bibitem{Grumiller:2008qz}
D.~Grumiller and N.~Johansson, ``{Instability in cosmological topologically
  massive gravity at the chiral point},''
  \href{http://dx.doi.org/10.1088/1126-6708/2008/07/134}{{\em JHEP} {\bf 07}
  (2008)  134}, \href{http://arxiv.org/abs/0805.2610}{{\tt arXiv:0805.2610
  [hep-th]}}.

\bibitem{Giribet:2008bw}
G.~Giribet, M.~Kleban, and M.~Porrati, ``{Topologically Massive Gravity at the
  Chiral Point is Not Chiral},''
  \href{http://dx.doi.org/10.1088/1126-6708/2008/10/045}{{\em JHEP} {\bf 10}
  (2008)  045}, \href{http://arxiv.org/abs/0807.4703}{{\tt arXiv:0807.4703
  [hep-th]}}.

\bibitem{Grumiller:2013at}
D.~Grumiller, W.~Riedler, J.~Rosseel, and T.~Zojer, ``{Holographic applications
  of logarithmic conformal field theories},''
  \href{http://dx.doi.org/10.1088/1751-8113/46/49/494002}{{\em J. Phys. A} {\bf
  46} (2013)  494002}, \href{http://arxiv.org/abs/1302.0280}{{\tt
  arXiv:1302.0280 [hep-th]}}.
  
\end{thebibliography}
\end{document}